\documentclass[a4paper,onecolumn,11pt,accepted=2024-12-06]{quantumarticle}
\pdfoutput=1
\usepackage{graphicx}
\usepackage{amsmath}
\usepackage{amssymb}
\usepackage{orcidlink}
\usepackage{hyperref}
\usepackage{multicol}
\newenvironment{Figure}
  {\par\medskip\noindent\minipage{\linewidth}}
  {\endminipage\par\medskip}

\newcommand{\rotop}{\hat{T}}  
\makeatletter
\newcommand{\lrarrow}{\mathrel{\mathpalette\lrarrow@\relax}}
\newcommand{\lrarrow@}[2]{%
  \vcenter{\hbox{\ooalign{%
    $\m@th#1\mkern6mu\rightarrow$\cr
    \noalign{\vskip3pt}
    $\m@th#1\leftarrow\mkern6mu$\cr
  }}}%
}

\let\origcontentsline\addcontentsline
\newcommand\stoptoc{\let\addcontentsline\nocontentsline}
\newcommand\resumetoc{\let\addcontentsline\origcontentsline}

\makeatother

\begin{document}
\title{Strategies for implementing quantum error correction in molecular rotation}

\author{Brandon J. Furey \orcidlink{0000-0001-7535-1874}}
\email{brandon.furey@uibk.ac.at}
\affiliation{Institut f\"ur Experimentalphysik, Universit\"at Innsbruck, Technikerstraße 25/4, 6020 Innsbruck, Austria}

\author{Zhenlin Wu
\orcidlink{0000-0002-8188-7701}}
\affiliation{Institut f\"ur Experimentalphysik, Universit\"at Innsbruck, Technikerstraße 25/4, 6020 Innsbruck, Austria}

\author{Mariano Isaza-Monsalve
\orcidlink{0009-0006-6996-7916}}
\affiliation{Institut f\"ur Experimentalphysik, Universit\"at Innsbruck, Technikerstraße 25/4, 6020 Innsbruck, Austria}
\author{Stefan Walser \orcidlink{0009-0001-0217-5117}}
\affiliation{Institut f\"ur Experimentalphysik, Universit\"at Innsbruck, Technikerstraße 25/4, 6020 Innsbruck, Austria}

\author{Elyas Mattivi
\orcidlink{0009-0008-9681-6667}}
\affiliation{Institut f\"ur Experimentalphysik, Universit\"at Innsbruck, Technikerstraße 25/4, 6020 Innsbruck, Austria}

\author{René Nardi \orcidlink{0009-0002-7533-6126}}
\affiliation{Institut f\"ur Experimentalphysik, Universit\"at Innsbruck, Technikerstraße 25/4, 6020 Innsbruck, Austria}

\author{Philipp Schindler \orcidlink{0000-0002-9461-9650}}
\email{philipp.schindler@uibk.ac.at}
\affiliation{Institut f\"ur Experimentalphysik, Universit\"at Innsbruck, Technikerstraße 25/4, 6020 Innsbruck, Austria}

\begin{abstract}
    The rotation of trapped molecules offers a promising platform for quantum technologies and quantum information processing. 
    In parallel, quantum error correction codes that can protect quantum information encoded in rotational states of a single molecule have been developed.
   These codes are currently an abstract concept, as no implementation strategy is yet known.
   Here, we present a step towards experimental implementation of one family of such codes, namely absorption-emission codes. We first construct architecture-agnostic check and correction operators.
   These operators are then decomposed into elements of the quantum logic spectroscopy toolbox that is available for molecular ions. 
   We then describe and analyze a measurement-based sequential as well as an autonomous implementation strategy in the presence of thermal background radiation, a major noise source for rotation in polar molecules. 
   The presented strategies and methods might enable robust sensing or even fault-tolerant quantum computing using the rotation of individual molecules.
\end{abstract}

\maketitle
\section{\label{sec:intro}Introduction}

Quantum computers promise to solve certain computational tasks more efficiently than existing classical computers, but the performance of current prototypes is limited by noise~\cite{Preskill2018}.
It is thus expected that quantum error correction (QEC) will be required for useful large-scale quantum information processors~\cite{Roffe2019}. While the ultimate purpose of QEC protocols is to reduce errors during the execution of quantum algorithms, they can also extend the storage time of quantum memories~\cite{RevModPhys.87.307}. Over the last decade, proof-of-principle QEC implementations have been demonstrated in multiple architectures~\cite{postler2023demonstration,krinner2022realizing, zhao2022realization, google2023suppressing,Bluvstein2023,putterman2024hardwareefficientquantumerrorcorrection}, culminating most recently in several demonstrations of operation beyond break-even in different hardware architectures~\cite{Sivak2023,Acharya2024}.

QEC protects quantum information by storing it redundantly, either over multiple physical information carriers, such as atomic ions or superconducting qubits~\cite{Roffe2019}, or in the complex Hilbert space of a single physical information carrier~\cite{PhysRevA.64.012310, BRADY2024100496,PRXQuantum.3.010335, albert2022bosonic,Mirrahimi_2014}. Here we concentrate on the latter approach, which has been used successfully in quantum harmonic oscillators. This is known as \emph{bosonic} quantum error correction and has been implemented in superconducting~\cite{Ofek2016} as well as trapped ion systems~\cite{Fluhmann2019}. In the former, an extension of the storage time of a superconducting quantum memory coupled to a microwave cavity has been demonstrated~\cite{Sivak2023,Ofek2016}. Furthermore, bosonic codes can also be embedded in finite dimensional Hilbert spaces, known as qudits~\cite{PhysRevA.64.012310}. There, an implementation of QEC in the spin qudits of molecules has been proposed~\cite{Chizzini2022, YUNGERHALPERN201992}.

Trapped molecules have recently emerged as a viable platform for quantum technologies by demonstrating entanglement~\cite{doi:10.1126/science.adf4272,doi:10.1126/science.adf8999} and relatively long quantum information storage times~ \cite{PhysRevLett.127.123202, doi:10.1126/science.aal5066}. Molecules feature multiple degrees of freedom (DOF) that can be used to store quantum information. We are interested in the rotational DOF of a single molecule which provides a large Hilbert space. The concept of bosonic QEC has been extended to molecular rotation in Ref.~\cite{PhysRevX.10.031050}. This framework has been used to develop \emph{absorption-emission codes} that protect against a major source of errors in small polar molecules: \emph{spontaneous photon emission} and \emph{interaction with black body radiation}~\cite{jain2023ae}.

In this work, we present and theoretically analyze implementation strategies for such codes in linear molecules using quantum logic spectroscopy (QLS) with co-trapped atomic ions~\cite{schmidt2005spectroscopy,Chou2017}. The Hamiltonian of the linear rotor is introduced, and absorption and emission errors in multiple regimes are discussed in Sec.~\ref{sec:intro:system}. We then revisit the conditions that need to be fulfilled to correct these errors in Sec.~\ref{sec:intro:errcond} and summarize the construction and properties of absorption-emission codes in Sec.~\ref{sec:intro:rotqec}. Furthermore, we describe a simplified variant of the code that is able to approximately correct for the same type of errors in Sec.~\ref{sec:intro:approxqec}.
We then develop architecture-agnostic check and correction operators to perform error correction that work for the original as well as the approximate codes in Sec.~\ref{sec:protocols}.
 
These operators are then decomposed into elements of the quantum logic spectroscopy toolbox that is available for molecular ions. We describe the quantum logic toolbox in Sec.~\ref{sec:implement:trappediontools} and analyze a measurement-based sequential implementation strategy in Sec.~\ref{sec:implement:exactqec}. Furthermore, we develop and analyze a measurement-free autonomous implementation strategy using dissipation engineering in Sec.~\ref{sec:implement:dec}. We discuss the requirements that the implementation imposes on molecular species and future research directions in Sec.~\ref{sec:conclusion}.

\subsection{\label{sec:intro:system}The linear rotor}

\begin{figure*}
    \centering
    \includegraphics[width=\textwidth]{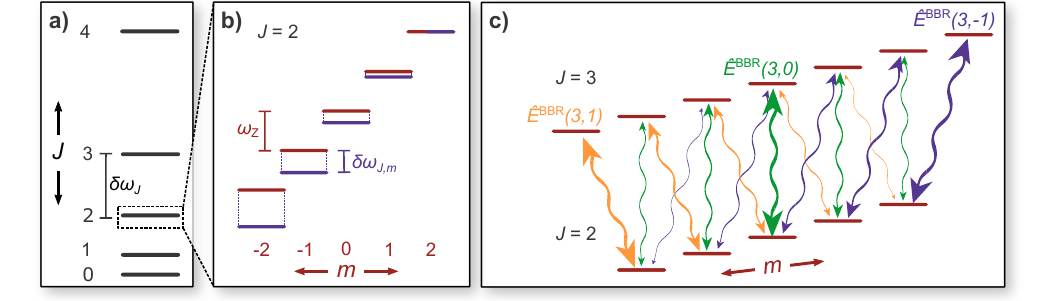}
    \caption{a) Sketch of the eigenfrequencies of the $\hat{J}^2$ term of $\hat{H}_{\textrm{rot}}$.  b) Nonlinear shifts ${\delta \omega_{J,m}}$, shown in violet, which can be induced in the Zeeman substructure which is shown in red. c) The unresolved ${\hat{E}^{\textrm{BBR}}(3,\delta m)}$ operators for interactions with blackbody radiation are shown here with the coupling strengths indicated by the thickness of the arrows. The operators are labeled by color above the sublevels at which their coupling strength is strongest.}
    \label{fig:intro_physical_system}
\end{figure*}

We consider a molecule modeled as a quantum mechanical linear rigid rotor with an electric dipole moment. The electric dipole moment gives rise to a magnetic dipole moment ${g_J \mu_{\textrm{N}} \hat{J} / \hbar}$ under rotation, and in the presence of a magnetic field ${\vec{B} = B \hat{z}}$ generates a Zeeman splitting by a ${\hat{J} \cdot \vec{B}}$ interaction. Such a system has energy eigenvalues and a Hilbert space described by the solution of the time-independent Schrödinger equation for the rotational Hamiltonian,
\begin{equation}
    \label{eq:hrotschrodinger}
    \hat{H}_{\textrm{rot}} = \frac{B_{\textrm{R}}}{\hbar^2} \hat{J}^2 + \frac{g_J\, \mu_{\textrm{N}}\, B}{\hbar} \hat{J}_z\, ,
\end{equation}
with the angular momentum operator $\hat{J}$ and its projection on the magnetic field quantization axis in the lab frame $\hat{J}_z$, where ${B_{\textrm{R}} = \hbar^2/(2I)}$ is the rotational constant for a rotor with moment of inertia $I$. The Zeeman splitting is given by ${\omega_Z = g_J\mu_{\textrm{N}} B/\hbar}$ with $g_J$ the rotational $g$-factor and $\mu_{\textrm{N}}$ the nuclear magneton. We define ${\omega_J = B_R\, J (J + 1) / \hbar}$ and ${\omega_{J,m} = \omega_J + m \omega_Z}$,
as indicated in Fig.~\ref{fig:intro_physical_system}. In this model, we neglect any other angular momentum couplings.
 
The eigenbasis of the Hamiltonian in Eq.~\eqref{eq:hrotschrodinger} can be described by the quantum numbers ${J,m}$ which denote the rotational quantum number and its projection on the quantization axis given by the magnetic field, respectively.
The rotational Hilbert space $\mathcal{H}$, neglecting any other contributions to angular momentum and any other DOF, is then the Hilbert space over the basis set ${\{|J,m\rangle\, :\ J\geq 0,\ m \in [-J,J],\ J,m \in \mathbb{Z}\}}$ . 

In the description of rotational transitions induced by coupling to the environment via interaction with blackbody radiation (BBR) or spontaneous decay (SD), or by coupling to radiation fields used for coherent control, we make extensive use of a \textit{rotational ladder operator} ${\rotop(J,m,\delta J, \delta m)}$, defined as
\begin{equation}
    \label{eq:shiftop}
    \rotop(J,m,\delta J, \delta m) = |J + \delta J, m + \delta m\rangle \langle J, m|\, .
\end{equation}
Direct transitions, \textit{i.e.}, single-photon electric dipole transitions between rotational manifolds, are described by this operator with ${|\delta J| = 1}$ and ${|\delta m| \in \{0,1\}}$. Raman transitions, utilized for control within $m$-sublevels, correspond to ${|\delta J| = 0}$ and ${|\delta m| \in \{1,2\}}$.

We aim to implement these transitions using quantum control methods that have been developed for quantum information processing with neutral and charged molecules~\cite{doi:10.1126/science.aal5066,doi:10.1126/science.aba3628}. These techniques require spectroscopic addressability of transitions between the individual states. This is certainly possible between $J$-manifolds due to their inherent anharmonicity. However, within a rotational manifold, the transitions between different $m$-sublevels are degenerate under the Hamiltonian $ \hat{H}_{\textrm{rot}}$. 

Thus, we require sufficient nonlinearity in the energy eigenvalues of the $m$-sublevels such that transitions between individual $m$-sublevels within a single $J$-manifold can be spectroscopically resolved. Such a nonlinearity can be inherent to the molecule, \textit{e.g.}, nuclear spin - rotation coupling~\cite{Chou2017}, or engineered, \textit{e.g.}, AC-Stark shifts which exhibit $m$-dependent coupling strengths~\cite{N_B_Delone_1999, Gregory2017}. 
 A nonlinear shift of each $m$-sublevel by frequency ${\delta \omega_{J,m}}$ is represented by the Hamiltonian 
\begin{equation}
    \hat{H}_{\delta} = \hbar \delta \omega_{J,m} |J,m\rangle \langle J,m |\, ,
\end{equation}
yielding an energy level structure such as shown in Fig.~\ref{fig:intro_physical_system}.

We denote the subspace of the Hilbert space in which we will encode the information as the codespace ${\mathcal{H}_{\textrm{C}} \subset \mathcal{H}}$, and in particular we require that this subspace exists in a single rotational manifold ${J_{\textrm{C}}}$. We define the minimum frequency difference between ${\delta J = 0}$ Raman transitions between $m$-sublevels in the code manifold as 
\begin{equation}
    \label{eq:nonlinearj0minmax}
    \delta \omega_{\textrm{Raman}}^{\min} = \underset{m \in [-J_{\textrm{C}},J_{\textrm{C}}]}{\min} \big\{|\delta \omega_{J_{\textrm{C}},m} - \delta \omega_{J_{\textrm{C}}, m'}|\big\}\, ,
\end{equation} 
where ${m' = m + \delta m}$ for ${\delta m \in \{\pm 1, \pm 2\}}$.
We also define the minimum and maximum frequency differences of ${|\delta J| = 1}$ direct transitions which couple to the code manifold as 
\begin{equation}
    \label{eq:nonlinearj1minmax}
    \delta \omega_{\textrm{direct}}^{\min,\max} = \underset{m \in [-J_{\textrm{C}},J_{\textrm{C}}]}{\min,\max} \big\{|\delta \omega_{J_{\textrm{C}},m} - \delta \omega_{J', m'}|\big\}\, ,
\end{equation}
where ${J' = J_{\textrm{C}} + \delta J}$ and ${m' = m + \delta m}$ for ${\delta m \in \{0, \pm 1\}}$.

We are interested in two different regimes of the nonlinear shifts: (i) the \textit{resolved} regime where the splitting ${\delta \omega_{J,m}}$ allows spectroscopic addressing of individual ${|J,m\rangle}$~states, \textit{i.e.}, the Rabi rate of driving the transition ${\Omega \ll \delta \omega_{\textrm{direct}}^{\min}}$ and (ii) the \textit{unresolved} regime where ideally ${\delta \omega_{J,m} = 0}$ and a single light field drives all $m$-sublevels resonantly. For nonzero ${\delta \omega_{J,m}}$, this regime is reached when the Rabi rate ${\Omega \gg \delta \omega_{\textrm{direct}}^{\max}}$. 

In the \textit{resolved} ${|\delta J| = 1}$ regime (i), each $m$-sublevel transition can be resolved. Applying a light field that is resonant with a specific transition ${\delta J, \delta m}$ results in the interaction Hamiltonian
\begin{equation}
    \label{eq:H_int_specific}
    \hat{H}^{\textrm{R}}_{\textrm{int}} = \hbar \Omega(J,m,\delta J, \delta m)    \big[\rotop(J,m,\delta J, \delta m) + \textrm{h.c.}\big] 
\end{equation}
where the coupling rate $\Omega$ is determined by the transition dipole moment and the electric field strength. Applying a rotating wave approximation allows us to neglect interactions with other rotational states. Incoherent interactions with the environment via electric dipole transitions can be modeled by a master equation with a single collapse operator per sublevel. In the case of stimulated emission or absorption of BBR, the dynamics between the states ${|J,m\rangle \leftrightarrow |J - 1,m + \delta m\rangle}$ are described by the Liouvillian operators
\begin{equation}
    \label{eq:bbrdoop_res}
    \hat{\mathcal{E}}^{\textrm{BBR}}(J,m,\delta m) =     \sqrt{\Gamma^{\textrm{BBR}}(J,m,\delta m)}\, \rotop(J,m,-1,\delta m) + \textrm{h.c.}
\end{equation}
which occur with rates
\begin{equation}
    \label{eq:resolved-bbr-rate}
    \Gamma^{\textrm{BBR}}(J,m,\delta m) =     \gamma^{\textrm{BBR}}_J \bigg(\frac{4 \pi}{3}\bigg) |c^J(J -1,m + \delta m, 1, \delta m)|^2\, ,
\end{equation}
where the ${m,\delta m}$-independent part of the rates, $\gamma_{J}^{\textrm{BBR}}$, are given in Appendix~\ref{sec:rottrans:einstein} and $c^{J}$, the Slater integrals, are defined in Appendix~\ref{sec:rottrans:matrixelements}.

In the \textit{unresolved} ${|\delta J| = 1}$ regime (ii), the nonlinearities in the transition frequencies are not resolved and we consider the transition frequency between all $m$-sublevels to be equal.
The interaction Hamiltonian with an external field on resonance with the transition is thus the sum over all resolved interaction Hamiltonians $\hat{H}^{\textrm{R}}_{\textrm{int}}$ in the connected $J$-manifolds,
\begin{equation}
    \label{eq:H_int_unresolved}
    \hat{H}^{\textrm{U}}_{\textrm{int}} = \sum_{m = -J}^J \hat{H}^{\textrm{R}}_{\textrm{int}}\, .
\end{equation}
The unresolved collapse operators describing interactions with BBR are 
\begin{equation}
    \label{eq:bbrdoop}
    \hat{E}^{\textrm{BBR}}(J,\delta m) = \sum_{m = -J}^J \hat{\mathcal{E}}^{\textrm{BBR}}(J,m,\delta m)\, . 
\end{equation}
These operators are shown schematically for ${J=3}$ in Fig.~\ref{fig:intro_physical_system}.

The set of errors which represent the couplings to the environment via unresolved absorption and emission includes not only interactions with BBR, but also spontaneous decay. The error model for these processes $\hat{E}^{\textrm{SD}}$ and their corresponding rates $\gamma_J^{\textrm{SD}}$ are given in Appendix~\ref{sec:rottrans:einstein}. The full set of unresolved errors thus comprises a family of collapse operators, 
\begin{equation}
    \label{eq:collapsefamily}
    \mathcal{C}_{\textrm{env}} = \big\{\hat{E}^{\textrm{SD}}(J,\delta m),\ \hat{E}^{\textrm{BBR}}(J,\delta m)    :\ J \in \mathbb{N},\ \delta m \in \{-1,0,1\}\big\}\, .
\end{equation}

\subsection{\label{sec:intro:errcond}Error correction conditions}

Protection of quantum information in the presence of absorption and emission processes requires an encoding of logical information such that the error operators defined above do not affect the encoded information. A set of errors, characterized by Kraus operators  $\hat{K}_a$, can be corrected if the Knill-Laflamme error correction conditions for a set of logical states~${|\overline{i}\rangle,\  |\overline{j}\rangle}$ are fulfilled~\cite{PhysRevA.55.900}:
\begin{equation}
    \label{eq:klqec}
    \langle \overline{i}|\hat{K}_a^\dagger \hat{K}_b |\overline{j}\rangle = c_{ab}\delta_{ij} \, .
    \end{equation}
The weights~$c_{ab}$ should be state-independent meaning that no information on the logical states should be gained by the environment, and the Kronecker delta~$\delta_{ij}$ is independent of the Kraus operators indicating that orthogonal logical states remain orthogonal for all error channels.

It should be noted that there exists no QEC code which can protect information in the resolved ${|\delta J| = 1}$ regime where the photons emitted or absorbed from each state are distinguishable~\cite{jain2023ae}. Any quantum superposition will be projected into a classical mixture after a single absorption or emission event. Thus, any implementation of QEC in the rotation of a molecule must be realized in the unresolved regime. 

In Ref.~\cite{jain2023ae}, it was shown how molecular states can be constructed to fulfill these conditions for errors corresponding to unresolved quantum jumps. These codes are referred to as absorption-emission~(Æ) codes. In particular, it was shown that the anharmonicity of the $J$-manifold centroid energies makes error correction impossible if the codewords span multiple $J$-manifolds. However, encoding quantum information in a single $J$-manifold can be correctable if the decay processes are degenerate across the $m$-sublevels, \textit{i.e.}, the processes act in the unresolved regime. Furthermore, in Ref.~\cite{jain2023ae} the argument was made that error correction is possible only if the frequency difference between absorbed or emitted photons, given by $\delta \omega_{\textrm{direct}}^{\max}$, is smaller than the natural linewidth of transitions out of the codespace so that the decay processes do not gain any information of the logical state. The linewidth for decays out of a given $J$-manifold is
\begin{equation}
    \label{eq:total-decay-linewidth-logical}
    \Gamma_J = \frac{1}{2 J + 1}
    \Big[ J (\gamma_{J}^{\textrm{SD}} + \gamma_{J}^{\textrm{BBR}}) + (J + 1)\,  \gamma_{J + 1}^{\textrm{BBR}} \Big] \, ,
\end{equation}
and the linewidth for decays out of the codespace is thus ${\Gamma_{\textrm{C}} = \Gamma_{J_{\textrm{C}}}}$. 

Alternatively, the linewidth of the absorbed or emitted photons can be controlled by  the rate at which these errors are measured. This can be accomplished by performing a check operation capable of detecting a ${|\delta J| = 1}$ jump out of the {$J_{\textrm{C}}$-manifold} without disturbing the logical state. In order to satisfy the error correction conditions, the rate of this projection process $R_{\textrm{proj}}$ needs to fulfill ${R_{\textrm{proj}} \gg \max \{\delta \omega_{\textrm{direct}}^{\max}, \Gamma_{\mathrm{C}}\}}$. This process can be interpreted as effectively broadening the sublevels of the error manifolds and ensures that errors can be projected before a second error occurs. The projection rate ${R_{\textrm{proj}} \gg \delta \omega_{\textrm{direct}}^{\max}}$ yields a projection into an undisturbed state or one where an unresolved quantum jump error has occurred. With the additional condition ${\delta \omega_{\textrm{Raman}}^{\min} \gg \Gamma_{\textrm{C}}}$, we can ensure that $m$-shifts can be corrected fast enough such that the probability for another absorption or emission event during this process is low.
Thus, we can establish the following hierarchy of timescales
\begin{equation}
    \label{eq:timescales}
        R_{\textrm{proj}} \gg \delta \omega_{\textrm{direct}}^{\max}\quad \text{and}\quad \delta \omega_{\textrm{Raman}}^{\min} \gg  \Gamma_{\textrm{C}} \, .
\end{equation}

In order to protect the logical quantum information, this hierarchy of timescales needs to be fulfilled at all times. It should be noted that $\delta \omega_{\textrm{Raman}}^{\min}$ sets the rate for addressing individual $m$-sublevels. 
Explicitly, this limits the full error correction rate $R_{\textrm{QEC}}$ to
\begin{equation}
    \label{eq:rprojrcorr}
    R_{\textrm{proj}}  \gg \delta \omega_{\textrm{Raman}}^{\min} \gg R_{\textrm{QEC}} \gg \Gamma_{\textrm{C}} \, .
\end{equation}
    
\subsection{\label{sec:intro:rotqec}Rotational Æ error correction codes and their properties}

We will here revisit the QEC conditions for the Æ-codes. It has been shown in Ref.~\cite{jain2023ae} that if the distance between populated $m$-sublevels is three or larger, only the symmetric  error correction criteria need to be fulfilled: 
\begin{equation}
    \label{eq:symmetricqeccond}
    \langle \overline{0}|\hat{K}_a^{\dagger} \hat{K}_b |\overline{0}\rangle = \langle \overline{1}|\hat{K}_a^{\dagger} \hat{K}_b |\overline{1}\rangle \, ,
\end{equation}
for any error process $\hat{K}_a$ in the set of unresolved Kraus operators corresponding to the set of recoverable collapse operators $\mathcal{C}_{\textrm{rec}}$. This subset is defined as the subset of $\mathcal{C}_{\textrm{env}}$ with a single absorption or emission collapse operator coupling from ${J_{\textrm{C}} \rightarrow J_{\textrm{C}} \pm 1}$ which in principle is recoverable by implementation of a first-order rotational error correction protocol:  
\begin{equation}
    \label{eq:corrsubset}
    \mathcal{C}_{\textrm{rec}} = \big\{ \hat{E}^{\textrm{SD}}(J_{\mathrm{C}},\delta m),\  \hat{E}^{\textrm{BBR}}(J_{\mathrm{C}},\delta m),    \hat{E}^{\textrm{BBR}}(J_{\mathrm{C}} + 1,\delta m)\ :\ \delta m \in \{-1,0,1\} \big\}\, .
\end{equation}

Satisfying Eq.~\eqref{eq:symmetricqeccond} requires that the probability to absorb or emit a photon with defined frequency and polarization to be the same for each of the logical basis states and thus no information on the logical state can leak to the environment.

\begin{figure}
    \centering
    \includegraphics[width=86.78 mm]{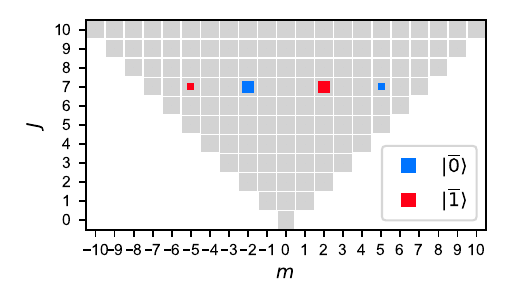}
    \caption{Codeword populations of the CS(7,2,5) code: ${|\overline{0}\rangle}$~(blue) and ${|\overline{1}\rangle}$~(red). Areas are proportional to the norm squared of the amplitudes.}
    \label{fig:basis}
\end{figure}

We focus on the counter-symmetric Æ codes that have been introduced in Ref.~\cite{jain2023ae}, where the encoded basis states are defined as 
\begin{align}
    \label{eq:cs0}
    |\overline{0}\rangle &= \sqrt{\frac{m_2}{M}} |J_{\textrm{C}},-m_1 \rangle + \sqrt{\frac{m_1}{M}} |J_{\textrm{C}},m_2 \rangle\\
    \label{eq:cs1}
    |\overline{1}\rangle &= \sqrt{\frac{m_1}{M}}  |J_{\textrm{C}},-m_2 \rangle + \sqrt{\frac{m_2}{M}} |J_{\textrm{C}},m_1 \rangle \, ,
\end{align}
where ${M = m_1 + m_2}$. This code family is identified as {CS($J_{\textrm{C}},m_1,m_2$)}, and the CS(7,2,5) code is shown graphically in Fig.~\ref{fig:basis}.
The subspace spanned by the complex linear combinations of ${\{|\overline{0}\rangle,\ |\overline{1}\rangle\}}$ is defined as the logical space $\mathcal{H}_{\textrm{C}}$.
This family of codes has been shown to satisfy the error correction criteria for the unresolved single  absorption or emission errors corresponding to $\mathcal{C}_{\textrm{rec}}$ when ${m_1 \geq 3/2}$, ${m_2 \geq m_1 + 3}$, and thus ${J_{\textrm{C}} \geq m_2 \geq 9/2}$. The counter-symmetric Æ codes can be encoded in either half-integer or integer angular momentum systems.

It is then straightforward to define the logical operators as
\begin{align}
    \label{eq:log-fid-ops}
        \overline{X} &= |J_{\textrm{C}}, -m_2 \rangle \langle J_{\textrm{C}}, m_2 | + |J_{\textrm{C}}, m_1 \rangle \langle J_{\textrm{C}}, - m_1 | + \textrm{h.c.}\\
    \begin{split}
        \overline{Z} &= |J_{\textrm{C}}, -m_1 \rangle \langle J_{\textrm{C}}, - m_1 | + |J_{\textrm{C}}, m_2 \rangle \langle J_{\textrm{C}}, m_2 |\\
        &\quad - |J_{\textrm{C}}, -m_2 \rangle \langle J_{\textrm{C}}, - m_2 | - |J_{\textrm{C}}, m_1 \rangle \langle J_{\textrm{C}}, m_1 |
    \end{split}
\end{align}
which introduces a natural way of characterizing the performance of QEC by the logical fidelities for the evolution of a logical state~\cite{Postler2022}
    \begin{align}
    \label{eq:logical-fidelity-z}
    \mathcal{F}_{0,1} &= \frac{1 \pm \langle \overline{Z} \rangle}{2}\\
    \mathcal{F}_{\pm} &= \frac{1 \pm \langle \overline{X} \rangle}{2} \, .    
    \end{align}

We now investigate the performance of this code under the following correction strategy: After a decay indicated by ${\delta J}$ and ${\delta m}$, the population of the decayed states in ${|J_{\textrm{C}}+\delta J,m+\delta m \rangle}$ is coherently brought back to its origin ${|J_{\textrm{C}},m \rangle}$ states. This operation preserves the orthogonality between the states ${|\overline{0}\rangle}$ and ${|\overline{1}\rangle}$ because the codewords never overlap in the recoverable error subspace.
On the other hand, an error alters the relative amplitudes of the ${|J_{\textrm{C}},m \rangle}$ states that encode the logical qubit according to the corresponding Slater integrals. Thus, the state after a single {$\delta J, \delta m$-correction} does not fulfill the error correction criteria anymore. The probability of another decay differs for the logical ${|\overline{0}\rangle,\ |\overline{1}\rangle}$~states. This effect can be directly quantified by a decrease in the logical fidelities of the logical {$\overline{X}$-basis} states ${|\overline{+}\rangle,\ |\overline{-}\rangle}$ after a second absorption or emission event. 

\subsection{\label{sec:intro:approxqec}Approximate Æ codes}

The error correction criteria can also be relaxed giving rise to ``approximate'' error correction codes~\cite{PhysRevA.56.2567, Zheng2023} which reduce the complexity to detect single errors. These approximate codes are characterized by relaxing the error correction criteria given by Eq.~\eqref{eq:symmetricqeccond} to
\begin{equation}
    \label{eq:relaxedqeccond}
    \langle \tilde{0}|\hat{K}_a^{\dagger} \hat{K}_b |\tilde{0}\rangle \approx \, \langle \tilde{1}|\hat{K}_a^{\dagger} \hat{K}_b |\tilde{1}\rangle\, ,
\end{equation}
which discards the condition that both logical basis states have the same error probabilities. In this case, an error event can partially project the logical information, causing a logical error.

One can thus define approximate protected codewords by simply directly encoding in the rotational states, \textit{e.g.}, 
\begin{align}
    |\tilde{0}\rangle &=  |J_{\textrm{C}}, m_{\tilde{0}}\rangle \\
    |\tilde{1}\rangle &=  |J_{\textrm{C}}, m_{\tilde{1}}\rangle\, ,
\end{align}
where the distance in $m$ between these codewords should satisfy ${|m_{\tilde{0}} - m_{\tilde{1}}| \geq 3}$ such that an absorption or emission event preserves orthogonality between basis states. But any ${|\delta m|=1}$ decay occurs with different probability for  ${|\tilde{0}\rangle}$ and ${|\tilde{1}\rangle}$ and thus information about the logical state can leak to the environment. This family of codes is identified as {A($J_{\textrm{C}},m_{\tilde{0}},m_{\tilde{1}}$)}.

\begin{figure}
    \centering
    \includegraphics[width=86.78 mm]{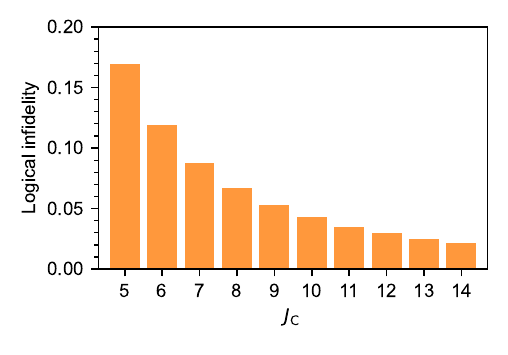}
    \caption{Worst-case logical infidelity after a single error for the approximate {A($J_{\textrm{C}}$,-2,2)} quantum error correction code as a function of the code manifold $J_{\textrm{C}}$.}
    \label{fig:approximate_J}
\end{figure}

The distinguishability of the codewords and thus the leakage of logical information to the environment will vanish at larger $J_{\textrm{C}}$. 
This effect is quantified by the worst-case logical infidelity ${1 - \mathcal{F}_+}$ over all single error events, shown as a function of $J_{\textrm{C}}$  for the codes {A($J_{\textrm{C}}$,-2,2)} in Fig.~\ref{fig:approximate_J}.

\section{\label{sec:protocols}Quantum error correction protocols}

We describe an error detection and correction strategy for a logical qubit encoded in a counter-symmetric Æ~code for ${J_{\textrm{C}} = 7}$, ${m_1 = 2}$, and ${m_2 = 5}$, referred to here as the CS(7,2,5) code. We also consider the ${|\tilde{0}\rangle = |7,-2\rangle}$, ${|\tilde{1}\rangle = |7,2\rangle}$ approximate code, identified as A(7,-2,2), using ideal check operators and unitary operations.

\begin{figure}
    \centering
    \includegraphics[width=86.78 mm]{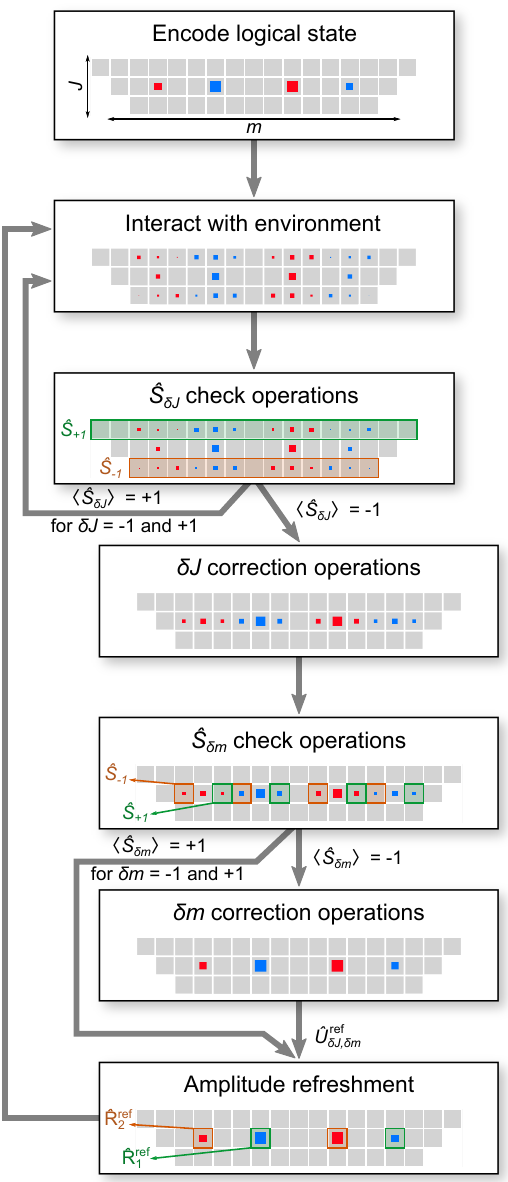}
    \caption{Algorithm for sequential quantum error correction of Æ codes in a linear rotor. The {$\hat{S}_{\delta J}$ check} and {$\delta J$-correction} operations are performed repeatedly and rapidly, and the subsequent {$\hat{S}_{\delta m}$ check}, {$\delta m$-correction}, and amplitude refreshment steps are performed only if a ${|\delta J| = 1}$ error is detected. Population of the CS(7,2,5) codewords and their evolution throughout the algorithm are indicated with red and blue squares. States which are acted on by operators are highlighted.}
    \label{fig:seqsketch}
\end{figure}

In our implementation strategy, we use spectroscopic resolvability of the {$m$-sublevels} for the correction of errors. In order to prevent interactions with the environment from also resolving the {$m$-sublevels}, we split up the correction of errors into three steps as shown in Fig.~\ref{fig:seqsketch}: 
\begin{enumerate}
    \item $\delta J$-detection and correction,
    \item $\delta m$-detection and correction,
    \item amplitude refreshment.
\end{enumerate}
This allows the {$\delta J$-correction} to be performed in the unresolved regime. Then, only if an error is detected, the resolved $\delta m$-correction operations are applied. 
This can be realized via two approaches: (a) setting the nonlinearity ${\delta \omega_{J,m} = 0}$ during ${\delta J}$ check and corrections, and only if an error is detected, applying the nonlinearity in the Zeeman sublevels followed by the {$\delta m$-correction} operations, or (b) the {$\delta J$-corrections} can be applied rapidly so as to not resolve the {$m$-sublevels} while ${\delta \omega_{J,m} \neq 0}$.

Detecting whether an absorption or emission event has occurred can be achieved with the measurement of two independent check operators $\hat{S}_{\delta J}$ for ${\delta J \in \{-1,1\}}$ identifying a population transfer out of the codespace in the {$J_{\textrm{C}}$-manifold}. 
The respective {$\hat{S}_{\delta J}$ operator} is defined as the diagonal operator ${\hat{S}_{\delta J} = \sum_{J,m} \lambda_{\delta J} |J,m\rangle \langle J,m|}$ with
\begin{equation}
    \label{eq:j-check-cases}
    \lambda_{\delta J} = 
    \begin{cases}
        -1& \text{if } J = J_{\textrm{C}} + \delta J\\
        +1 & \text{otherwise}\, .
    \end{cases}
\end{equation}

The expectation value of both operators $\hat{S}_{\delta J}$ are measured subsequently, where the outcome $-1$ indicates a population transfer into the ${J_{\textrm{C}} + \delta J}$ manifold. If either of these check operations indicate an absorption or emission event, the rotational state needs to be returned to the code manifold $J_{\textrm{C}}$. This can be achieved by conditionally applying the correction operator 
\begin{equation}
    \label{eq:correctjop}
    \hat{U}_{\delta J} = \sum_m \rotop(J_{\textrm{C}} + \delta J, m, -\delta J, 0) + \textrm{h.c.},
\end{equation}
acting identically on all $m$-sublevels.

An absorption or emission event also alters the projection of the angular momentum on the quantization axis by $\delta m \in \{-1,0,1\}$, depending on the polarization of the absorbed or emitted photon. This change in $m$ spreads the logical qubit into different {$m$-sublevels} within the {$J_{\textrm{C}}$-manifold} after applying the {$\delta J$-correction}, and thus the encoded information is not completely restored to the codespace. However, the encoded information remains in the recoverable error subspace. We refer to the effective shift errors in $m$ caused by sigma transitions followed by {$J$-correction} as Zeeman errors. 
This shift can be measured by a second pair of check operators $\hat{S}_{\delta m}$ which need to be defined such that no information on the encoded information can be gained. The diagonal $\delta m$-check operators ${\hat{S}_{\delta m} = \sum_{J,m} \lambda_{\delta m} |J,m\rangle \langle J,m|}$ are applied after the {$\delta J$-correction} and it is thus sufficient to define them in the {$J_{\textrm{C}}$-manifold}: 
\begin{equation}
    \label{eq:m-check-cases}
    \lambda_{\delta m} = 
    \begin{cases}
        -1& \text{if } |J, m - \delta m\rangle \in \mathcal{H}_{\textrm{C}} \\
        +1 & \text{otherwise},
    \end{cases}
\end{equation}
where the $m$-sublevels in $\mathcal{H}_{\textrm{C}}$ are $\{ -5, -2, 2, 5\}$ for the exact CS(7,2,5) code and $\{-2,2\}$ for the approximate code.
Again, a correction needs to be applied for an outcome of $-1$. The correction is performed using the conditional application of 
\begin{equation}
    \label{eq:correctmop}
    \hat{U}_{\delta m} = \sum_m \rotop(J_{\textrm{C}}, m, 0, -\delta m) + \textrm{h.c.}
\end{equation}

Addressing individual $m$-sublevels breaks the hierarchy of timescales for $R_{\textrm{proj}}$ as shown in Eq.~\eqref{eq:timescales} and should thus only be performed if an absorption or detection event has been detected. These operations have a limited rate, but this can still be performed at ${R_{\textrm{QEC}} \gg \Gamma_{\textrm{C}}}$ so the rate at which a second error occurs during the correction cycle can be kept comparatively low. Additionally, these operations are required relatively seldom, namely only if an absorption or emission occurs.

An error alters the relative amplitudes of the ${|J_{\textrm{C}},m \rangle}$~states that encode the logical qubit according to the corresponding Slater integrals, and this persists after the equal coupling correction of $J$ and $m$. 
In the exact counter-symmetric codes, this can be expressed as a coupling between the states ${|J_{\textrm{C}}, -m_1\rangle \leftrightarrow |J_{\textrm{C}}, m_2\rangle}$,  and ${|J_{\textrm{C}}, -m_2\rangle \leftrightarrow |J_{\textrm{C}}, m_1\rangle}$ respectively.
This coupling is an effective unitary operation comprising two rotations between the states in these two pairs of sublevels,
\begin{equation}
    \label{eq:unitary-refresh-err}
    \hat{U}_Q (\theta_{\delta J, \delta m, 1}, \theta_{\delta J, \delta m, 2}) =     \prod_{k=1}^2 \exp (-i \theta_{\delta J,\delta m,k}\,  \hat{R}_k / 2)\, ,
\end{equation}
where
\begin{equation}
    \label{eq:refresh-err}
    \hat{R}_k =  i \rotop(J_{\textrm{C}},-m_k,0,m_1+m_2) + \textrm{h.c.},
\end{equation}
for ${k\in \{1,2\}}$. The rotation angles $\theta_{\delta J,\delta m, k}$ are determined by the $\delta J$ and $\delta m$ of the error and are derived in Appendix~\ref{sec:append:sequentialqec}.

In the counter-symmetric codes, the amplitudes of the logical state can be refreshed after an error event and correction cycle once ${\delta J}$ and ${\delta m}$ are known. This can be achieved by coupling the states ${|J_{\textrm{C}}, -m_1\rangle}$ to ${|J_{\textrm{C}}, m_2\rangle}$; and ${|J_{\textrm{C}}, -m_2\rangle}$ to ${|J_{\textrm{C}}, m_1\rangle}$, respectively, using the inverse of the unitary operation in Eq.~\eqref{eq:unitary-refresh-err}, which is just the same rotation operators but with opposite rotation angles, \textit{i.e.}, 
\begin{equation}
    \label{eq:unitary-refresh}
    \hat{U}_{\textrm{ref}}^{\delta J, \delta m} = \hat{U}_Q(-\theta_{\delta J, \delta m,1},-\theta_{\delta J, \delta m, 2})\, .
\end{equation}

Refreshment of a corrected ${\delta J = +1}$, ${\delta m = -1}$ transition in the CS(7,2,5) code can be performed with angles ${-\theta_{1,-1,1} = -0.7160}$ and ${-\theta_{1,-1,2} = 0.7145}$ rad. Consider the initial state ${|\overline{+}\rangle}$ in this code experiencing this decay: The logical fidelity after {${\delta J,\delta m}$-correction} is reduced to ${\mathcal{F}_+ = 0.877}$. With amplitude refreshment, this fidelity can be restored to ${\mathcal{F}_+ = 1}$. This operation cannot be performed on the approximate codes, where the change in amplitudes manifests as a logical error via a rotation in the logical Bloch sphere.

\section{\label{sec:implement}Implementation strategies for trapped ions}

\subsection{\label{sec:implement:trappediontools}Trapped molecular ion toolbox}

We describe and analyze an implementation strategy for the error detection and correction operations described above in an ion trap platform where a single molecular ion is co-trapped with an atomic ion~\cite{schmidt2005spectroscopy,Chou2017,doi:10.1126/science.aba3628,Wolf2016,doi:10.1126/science.aaz9837,Lin2020}. In such a system the combined motional state of the molecule and the atom can serve as a reliable bus to transfer quantum information between the atom and the molecule~\cite{PhysRevLett.75.4011}. This serves as the basis for the quantum logic spectroscopy toolbox, enabling high-fidelity state initialization and readout of the molecule via the co-trapped atom~\cite{schmidt2005spectroscopy}. 

In order to keep the number of parameters of the system small, we consider only a single error channel: blackbody radiation (BBR) absorption and stimulated emission on the molecular rotation. Assuming perfect operations on the atomic system, it is sufficient to model the ion trap and the atom as a single motional mode which significantly reduces the required resources for numerical analysis. Thus, we consider the Hamiltonian
\begin{equation}
    \label{eq:hamilrotmotatom}
    \hat{H}_{\textrm{tot}} = \hat{H}_\textrm{rot} \otimes \hat{H}_\textrm{motion},
\end{equation}
with ${\hat{H}_\textrm{motion} = \sum_i \hbar \omega_i (1/2 + \hat{a}_i^\dagger \hat{a}_i)}$.

We model the manipulation of the rotational states by the interaction Hamiltonian 
\begin{equation}
    \label{eq:sidebandcarrierhamil}
    \hat{H}_{\textrm{int}}({J,m,\delta J, \delta m}) =
    \hbar \Omega(J,m,\delta J,\delta m)\, \rotop(J,m,\delta J,\delta m)
    \otimes \hat{M}_{\textrm{int}}    + \text{h.c.},
\end{equation}
where the operator $\hat{M}_{\textrm{int}}$ indicates the action that a rotational population transfer has on the motional state. Driving the molecular transition on resonance leaves the motional state unchanged and therefore ${\hat{M}_{\textrm{int}}=\hat{M}_{\textrm{CAR}}=\hat{\mathbb{I}}}$. This operation is known as a carrier~(CAR) transition.
  
When the frequency of the interacting field differs from the frequency of the carrier transition by the frequency of the motional sideband, and the effective {$\vec{k}$-vector} has nonzero projection on the motional mode axis, the operator acquires the form ${\hat{M}_{\textrm{BSB}}=\hat{a}^\dagger}$. This adds a phonon to the motional state during a population transfer from the ground to the excited state and such an operation is known as a blue sideband~(BSB) transition. Conversely, a decrease in frequency from the carrier by the motional mode frequency can drive a red sideband~(RSB) transition with ${\hat{M}_{\textrm{RSB}}=\hat{a}}$~\cite{Wineland1998-up}. The coupling strength of a sideband transition, relative to that of a carrier transition, is reduced by the Lamb-Dicke factor~$\eta$, which characterizes the strength of the coupling between the internal state of the atomic or molecular ion and the motional state.

We consider here two types of interactions between light fields and molecular rotations with different selection rules: (i) Raman carrier and sideband transitions on a molecule with ${|\delta J| = 0}$ and ${|\delta m| \leq 2}$, and (ii) direct microwave/terahertz carrier and sideband transitions on a molecule with ${|\delta J|=1}$ and ${|\delta m| \leq 1}$.\footnote{We allow microwave sideband transitions to simplify the implementation. It might not be possible to drive sideband operations directly between rotational manifolds within the electronic and vibrational ground state in an experiment due to the small Lamb-Dicke factor. In that case, a combination of optical Raman ${|\delta J| = 2}$ sideband and microwave carrier ${|\delta J| = 1}$ transitions or a strong magnetic field gradient can be used~\cite{Ospelkaus2011}.} These transitions are used to control the rotational state, including check and correction operations. 

The motional state can then be read out on an atomic ion by attempting to drive a RSB $\pi$-pulse on the atomic qubit transition, performing fluorescence readout of the qubit, and recooling of the motional mode~\cite{schmidt2005spectroscopy,Wineland1998-up}.
In the following, we can  model the atomic system by a perfect measurement of the phonon number in the shared motional mode.

\subsection{\label{sec:implement:exactqec}Sequential check operators and correction}

\paragraph*{\label{sec:implement:exactqec:dj}Sequential $\delta J$ check and correction}

The {$J$-projection} and correction operation can be implemented by (i) mapping the information of a decay event in the molecular rotation onto the motional mode and then (ii) mapping the motional mode onto the co-trapped atomic ion electronic state and reading out this state. The motional sideband operation is implemented using the unitary operator
\begin{equation}
    \label{eq:motionalsidebandjop}
    \hat{U}^{\delta J}_{\textrm{BSB}} =     \prod_m \exp\bigg[\frac{-i}{2}  \hat{H}_{\textrm{BSB}}(J_{\textrm{C}} + \delta J,m,-\delta J,0)\, t_{\textrm{opt}}\bigg]
\end{equation}
for time $t_{\textrm{opt}}$ to drive an optimal population transfer between the $J_{\textrm{C}} + \delta J \rightarrow J_{\textrm{C}}$ manifolds.  

The operation is performed at the timescale of $R_{\textrm{proj}}$ and should stay in the unresolved regime in order to not distinguish different {$m$-sublevels}. The requirement on the Rabi frequency of the molecular transition is then ${\eta \Omega \approx R_{\textrm{proj}} \gg  \delta \omega_{\textrm{direct}}^{\max}}$.

Performing these sideband pulses with a single {$\pi$-transition} yields an {$m$-dependent} coupling strength in $\Omega$. 
This would degrade the error correction performance because it cannot detect the complete ${J_{\textrm{C}} \pm 1}$ population. For the CS(7,2,5) code, the physical fidelity of the state after {$J$-correction} with a single pulse compared to an ideal detection operation with equal coupling is 97.5\% in the case of a decay with ${\delta m = 0}$. 

The problem of performing quantum operations with high fidelity in the presence of varying coupling strength is ubiquitous in practical quantum information processing. For example, the SCROFULOUS composite pulse sequence has been developed to be less sensitive to variations in coupling strength~\cite{PhysRevA.67.042308}. When applying a SCROFULOUS sequence for the {$J$-correction}, the physical fidelity of the state after correction as compared to an ideal operation with equal coupling increases to 99.9\%.

\paragraph*{\label{sec:implement:exactqec:dm}Sequential $\delta m$ check and correction}

For the detection and correction of a shift in $m$ by ${|\delta m| = 1}$, we can employ a BSB Raman operation ${\hat{H}_{\textrm{BSB}}(J_{\textrm{C}},m,0,-\delta m)}$, for which the individual {$m$-sublevels} can be spectroscopically resolved.  Thus, the coupling strength can be controlled for each transition individually and no composite pulse sequence is required. 

The sideband operation needs to be implemented four times, mapping information from the states ${|J_{\textrm{C}},m_{\textrm{C}}\rangle}$ for ${m_{\textrm{C}} \in \{\pm m_1, \pm m_2\}}$ onto the motional mode resulting in the unitary operation
\begin{equation}
    \label{eeq:motionalsidebandmop}
    \hat{U}_{\textrm{BSB}}^{\delta m} =         \prod_{m_{\textrm{C}}} \exp\bigg[\frac{-i}{2} \hat{H}_{\textrm{BSB}}(J_{\textrm{C}},m_{\textrm{C}}^\prime ,0,-\delta m) t_{\pi}(m_{\textrm{C}},\delta m)\bigg]
\end{equation}
with pulse duration $t_{\pi}(m_{\textrm{C}},\delta m) = {\pi/ \Omega(J_{\textrm{C}},m_{\textrm{C}},0,-\delta m)}$ and the target {$m$-substate} $m_{\textrm{C}}^\prime = m_{\textrm{C}}+ \delta m$.

\paragraph*{\label{sec:implement:exactqec:refresh}Amplitude refreshment}In order to refresh the amplitudes of the codewords, populations need to be exchanged between the constituents of both logical basis states ${|\overline{0}\rangle}$ and ${|\overline{1}\rangle}$. These operations require a change in quantum numbers ${|\delta J| = 0},\ {|\delta m| = m_1 + m_2 = 7}$. Driving transitions with ${|\delta m|=7}$ is not possible due to selection rules and thus the operation needs to be split into a sequence of ${|\delta m|=2}$ and ${|\delta m|=1}$ Raman operations. For the state ${|\overline{0}\rangle}$, this sequence transfers populations from ${m=-5}$ to ${m=1}$, performs a ${\delta m = 1}$~Raman operation to realize the actual refreshment, and implements the inverse of the transfer Raman sequence. The individual operations are based on carrier operations 
\begin{equation}
    \label{eeq:motionacarmop}
    \hat{U}_{\textrm{CAR}}^{m_i, \delta m} =          \exp\bigg[\frac{-i}{2} \hat{H}_{\textrm{CAR}}(J_{\textrm{C}},m_i ,0,-\delta m)\, t(m_i)\bigg]\, .
\end{equation}

The operator that transfers the populations of the logical state ${|\overline{0}\rangle}$  to ${m=1}$  can be written as 
\begin{equation}
    \label{eeq:refreshtransfer}
     \hat{U}^0_\textrm{trans} = \prod_{m_i} \hat{U}_{\textrm{CAR}}^{m_i,\delta m} 
\end{equation}
with ${\delta m = 2}$, ${m_i \in \{-5,-3,-1\}}$, and ${t(m_i)}$ corresponding to a {$\pi$ rotation} on the respective transition. The refresh operation is given by 
\begin{equation}
    \label{eeq:refreshtdeltam1}
     \hat{U}^0_\textrm{ref} = \hat{U}_{\textrm{CAR}}^{m_i,\delta m} 
\end{equation}
with ${\delta m = 1}$, ${m_i = 1}$, and ${t(m_i)}$ the time to implement the refreshment angle given in Eq.~\eqref{eq:amp-refresh-angles-err}. The second transfer is then the inverse of the initial transfer operation. The full refreshment operator for the state ${|\overline{0}\rangle}$ can then be written as 
\begin{equation}
    \label{eeq:refreshlog0}
     \hat{U}_\textrm{total}^0 = (\hat{U}_\textrm{trans}^0)^{\dagger} \, \hat{U}_\textrm{ref}^0 \, \hat{U}_\textrm{trans}^0 \, .
\end{equation}
This procedure must also be performed for the logical state ${|\overline{1}\rangle}$ where the parameters of the  transfer operations are ${\delta m = 2}$, ${m_i \in \{-2,0,2\}}$ and ${m_i=4}$ for the refreshment operation. The procedure for both basis states requires 12 additional Raman {$\pi$-operations} in addition to the two refreshment operations. Each operation can be performed at a rate~${\sim \delta \omega_{\textrm{Raman}}^{\min}}$. 

\paragraph*{\label{sec:implement:exactqec:numsim}Numerical simulation}

\begin{figure}
    \centering\includegraphics[width=86.78 mm]{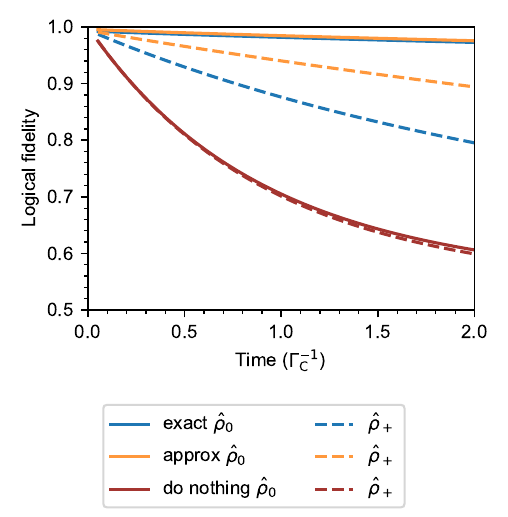}
    \caption{Sequential correction of the logical states~${|\overline{0}\rangle}$~{(solid line)} and ${|\overline{+}\rangle}$~(dashed line) using the exact {CS(7,2,5)}~(blue) and the approximate {A(7,-2,2)}~(orange) error correction code for the correction rate ${\Omega_\textrm{BSB}=500\, \Gamma_{\textrm{C}}}$.} 
    \label{fig:log_corr_cont}
\end{figure}

We analyze the performance of multiple correction cycles by numerical simulation using the QuTiP package~\cite{JOHANSSON20121760,JOHANSSON20131234}. 
We choose a noise model that focuses on the inherent capabilities of the code rather than on implementation details and thus the only noise source considered is absorption and stimulated emission by BBR on the molecular rotation during the operations. While performing the $\hat{U}_{\textrm{BSB}}^{\delta J}$ check and correction operations in the unresolved regime, the interaction with BBR is modelled by solving the Lindblad master equation with collapse operators as defined in Eq.~\eqref{eq:bbrdoop}. This noise model preserves superpositions between the individual $m$-sublevels. On the other hand, decays which occur during $m$-resolved operations of the $m$-correction and amplitude refreshment will destroy any superposition of the Zeeman sublevels in a rotational manifold and are thus more detrimental. The decay dynamics for this resolved regime is given by the Lindblad master equation using the operators given in Eq.~\eqref{eq:bbrdoop_res} in Appendix~\ref{sec:append:dec}.
To simplify the noise model, we assume that all operations can be performed at the same timescale, which is valid if the degeneracy of the $m$-levels can be lifted with a comparatively strong AC-Stark shift as discussed in Sec.~\ref{sec:intro:system}. 

The relevant timescale of the operations is then given by the Rabi frequencies of the sideband operations $\Omega_\textrm{BSB}$ in relation to the decay rate of the codespace. We assume a generic model with constant field strength of the thermal radiation, yielding ${\gamma_J^{\textrm{BBR}} = \gamma^{\textrm{BBR}}}$, independent of $J$. Then the {${J,m,\delta m}$-independent} part of the decay rate is just the total decay rate out of the code manifold
\begin{equation}
    \label{eq:standard-decay-rates}
    \gamma^{\textrm{BBR}} = \Gamma_{\textrm{C}}\, ,
\end{equation}
where ${\Gamma_{\textrm{C}} = 1}$ determines the timescale of the simulations.

We assume a quantum memory operation, where multiple rounds of error correction are applied, spaced by a determined free evolution time; in this example, ${0.05 \, \Gamma_{\textrm{C}}^{-1}}$. In Fig.~\ref{fig:log_corr_cont}, we show the logical fidelity with respect to the initial states~${|\overline{0}\rangle}$ and ${|\overline{+}\rangle}$ for the exact {CS(7,2,5)} and the approximate {A(7,-2,2)} code with correction timescale ${\Omega_\textrm{BSB}= 500\, \Gamma_{\textrm{C}}}$. The simulations show that the QEC protocol indeed extends the qubit storage time for the chosen parameters, although the initial state ${|\overline{+}\rangle}$ is more affected by noise.  We observe that the approximate code outperforms the exact code for this parameter regime. The approximate code requires less time for the {$\delta m$-correction} and does not have amplitude refreshment which incurs a lower penalty on the fidelity despite information on the logical state leaking to the environment. This improvement in fidelity of the operations dominates over the additional error that is introduced by the approximate nature of the code itself. 

\subsection{\label{sec:implement:dec}Autonomous approximate QEC using dissipation engineering}

While a sequential correction sequence which can project errors and measure their syndromes is in principle better suited to preserve coherences of the encoded logical state over long times due to the ability to refresh amplitudes, it can be difficult to implement in practice due to the large number of carefully-engineered pulses required. Here we consider a continuous approximate error correction protocol, referred to as dissipative error correction (DEC), which trades some loss of fidelity over time for less complex implementation. Conditions for autonomous error correction similar to the Knill-Laflamme conditions have been found in Ref.~\cite{Lihm2018}. Here we focus on approximate schemes, considering first a scheme that repumps molecular rotational states back to the $J_{\textrm{C}}$ manifold but without correcting errors that cause changes in $m$. Then we concatenate the implementation of a dissipative Zeeman correction scheme.

\paragraph*{\label{sec:implement:dec:repumping}Dissipative $\delta J$ correction}

The hierarchies of timescales described in Eqs.~\eqref{eq:timescales}~and~\eqref{eq:rprojrcorr} still need to be upheld. The dissipative scheme can be implemented by applying a continuous optical or microwave field with frequency and polarization chosen to drive a $\pi$-transition back to $J_{\textrm{C}}$ on a sideband of an additional DOF, \textit{e.g.}, a shared motional sideband in a system of co-trapped molecular and atomic ions. Two modes are required to cover the two possible directions for decays out of the $J_{\textrm{C}}$-manifold. Formally, the Hilbert space is modeled as the tensor product of the rotational space with a bimodal Fock space representing the motional state of the two modes: 
\begin{equation}
    \label{eq:dec-hilbert-repump}
    \mathcal{H} = \underbrace{|J,m\rangle}_{\text{rotation}} \otimes \underbrace{|n_{\downarrow}\rangle \otimes |n_{\uparrow}\rangle}_{\text{motion}}\, ,
\end{equation}
where the atomic state space is not considered as it is used for dissipation via cooling and thus contains no logical information. 

Such a dissipative interaction on the motional modes can be realized by sideband cooling of the co-trapped atomic ion. The irreversibility of dissipation to the environment results in a directional coupling from the ${J_{\textrm{C}} \pm 1}$ manifolds back to $J_{\textrm{C}}$. 
As time evolves, such a scheme should preserve ${\langle J \rangle = J_{\textrm{C}}}$ but results in a mixed state of different shifts in $m$ from the initial state.

This protocol is described by the Lindblad master equation, where the interaction Hamiltonian is given by a coherent blue sideband drive in the unresolved regime.
Thus, we can express the interaction Hamiltonian as a coupling strength modified by the Slater integral for the corresponding transition,
\begin{widetext}
\begin{multline}
    \label{eq:repump-hamiltonian}
    \hat{H}_{\textrm{DEC}}^{\delta\textrm{J}} = \sqrt{\frac{4 \pi}{3}}\hbar \sum_{m \in [-J_{\textrm{C}}, J_{\textrm{C}}]} \Big[  \Omega_{\downarrow} c^{J_{\textrm{C}} + 1}(J_{\textrm{C}},m,1,0)\, \rotop(J_{\textrm{C}} + 1,m,-1,0) \otimes \hat{a}_{\downarrow}^{\dagger} \otimes \hat{\mathbb{I}}_{\uparrow}\\
    + \Omega_{\uparrow} c^{J_{\textrm{C}} - 1}(J_{\textrm{C}},m,1,0)\, \rotop(J_{\textrm{C}} - 1,m,1,0) \otimes \hat{\mathbb{I}}_{\downarrow} \otimes \hat{a}_{\uparrow}^{\dagger} \Big] + \textrm{h.c.}
\end{multline}
where ${\sqrt{4 \pi / 3}\,  \Omega_{\downarrow,\uparrow}c^{J_{\textrm{C}} \pm 1}(J_{\textrm{C}},m,1,0)}$ are the {$m$-specific} coupling rates and ${\hat{a}^{\dagger}(\hat{a})}$ are the creation(annihilation) operators for the corresponding motional mode. The ${\downarrow,\uparrow}$ notation indicates in which direction the {${\delta J}$-correction} is performed. 
\end{widetext}

The interaction of the molecular rotational states with photons in the environment is modeled via collapse operators. These interactions include the effects of BBR and spontaneous decay. These operators are given by the family ${\{\hat{C} \otimes \hat{\mathbb{I}}_{\downarrow} \otimes \hat{\mathbb{I}}_{\uparrow}\,:\ \hat{C} \in \mathcal{C}_{\textrm{env}}\}}$
with ${\mathcal{C}_{\textrm{env}}}$ defined in Eq.~\eqref{eq:collapsefamily}. The dissipation of the motional modes, which can be accomplished via sideband cooling, is modeled by the collapse operators 
\begin{equation}
    \label{eq:cool-diss}
    \begin{split}
        \hat{C}_{\downarrow}^{\textrm{cool}} &= \sqrt{\Gamma_{ \downarrow}^{\textrm{cool}}}\  \hat{\mathbb{I}}_{J,m} \otimes \hat{a}_{\downarrow} \otimes \hat{\mathbb{I}}_{\uparrow}\\ 
        \hat{C}_{\uparrow}^{\textrm{cool}} &= \sqrt{\Gamma_{\uparrow}^{\textrm{cool}}}\  \hat{\mathbb{I}}_{J,m} \otimes \hat{\mathbb{I}}_{\downarrow} \otimes \hat{a}_{\uparrow},
    \end{split}
\end{equation}
where ${\Gamma_{\downarrow,\uparrow}^{\textrm{cool}} \sim \Omega_{\downarrow,\uparrow} \gg \Gamma_{\textrm{C}}}$. The evolutions of the expectation value ${\langle J \rangle}$ and the physical fidelity of the codewords are shown in Appendix~\ref{sec:append:dec:decj}. 

\paragraph*{\label{sec:implement:dec:zeeman}Full correction}

Although repumping a rotational manifold can stabilize population in that manifold, the logical fidelity is reduced due to the population spreading out among the Zeeman sublevels that are not in the codespace. 

A full DEC scheme for ${|\delta J| = 1}$ decays must also correct for these Zeeman errors, which require corrections of the form ${\rotop(J_{\textrm{C}},m_{\textrm{C}} \pm 1,0,\mp 1)}$. Thus the transitions must be sufficiently non-degenerate as to be able to apply frequency-selective ${\delta J = 0}$ Raman sigma transitions on the correctable error subspace. Given the need for frequency-resolvability to apply the required polarization for the correction, the rate of the correction is necessarily slow. 
Frequency-resolvability of the Raman transitions between Zeeman sublevels allows the coupling strengths for each transition to be independently controlled and thus equal coupling for all {$m$-sublevels} can be applied. 
In a counter-symmetric Æ~code, a {$\sigma_+$-correction} acts on the 4 states~${\big\{|J_{\textrm{C}}, m_{\textrm{C}} - 1\rangle\ : m_{\textrm{C}} \in \{\pm m_1, \pm m_2\} \big\}}$ simultaneously, while a {$\sigma_-$-correction} acts on the set ${\big\{|J_{\textrm{C}}, m_{\textrm{C}} + 1\rangle\ : m_{\textrm{C}} \in \{\pm m_1, \pm m_2\} \big\}}$.

We utilize separate motional modes for the dissipation in the correction scheme in order to avoid coherent population trapping~\cite{BDAgapev_1993}.
Utilizing two additional motional modes results in the promotion of the Hilbert space to include the tensor product of the rotational space with a four-mode Fock space: 
\begin{equation}
    \label{eq:full-dec-hilbertspace}
    \mathcal{H} = \underbrace{|J,m\rangle}_{\text{rotation}} \otimes \underbrace{|n_{\downarrow}\rangle \otimes |n_{\uparrow}\rangle \otimes |n_{\rightarrow}\rangle \otimes |n_{\leftarrow}\rangle}_{\text{motion}}\, ,
\end{equation}
where again the atomic state space is not considered as it is used for dissipation via cooling and thus contains no logical information.

We can model the Zeeman DEC with the Hamiltonian
\begin{equation}
    \label{eq:zeemancorrect-hamiltonian}
    \hat{H}_{\textrm{DEC}}^{\delta \textrm{m}} = \hbar \sum_{m_{\textrm{C}}}\\
    \Big[ \Omega_{\rightarrow} \, \rotop(J_{\textrm{C}},m_{\textrm{C}} - 1,0,1) \otimes \hat{a}_{\rightarrow}^{\dagger} \otimes \hat{\mathbb{I}}_{\leftarrow} + \Omega_{\leftarrow} \, \rotop(J_{\textrm{C}},m_{\textrm{C}} + 1,0,-1) \otimes \hat{\mathbb{I}}_{\rightarrow} \otimes \hat{a}_{\leftarrow}^{\dagger} \Big] + \textrm{h.c.}
\end{equation}
where $\Omega_{\lrarrow}$ are the Rabi rates and ${\hat{a}^{\dagger}(\hat{a})}$ are the creation(annihilation) operators for the corresponding motional mode. Here, the modes ${|n_{\downarrow}\rangle \otimes |n_{\uparrow}\rangle}$ are neglected for simplicity. The $\lrarrow$~notation indicates in which direction the {$\delta m$-correction} is performed.
Dissipation of these motional modes is again accomplished via sideband cooling, modeled by collapse operators similar to the ones defined in Eq.~\eqref{eq:cool-diss}. For details of the evolution of the physical fidelity under Zeeman DEC, refer to Appendix~\ref{sec:append:dec:decz}. 

The full DEC scheme is then realized by applying ${\hat{H}_{\textrm{DEC}}^{\delta \textrm{J}} + \hat{H}_{\textrm{DEC}}^{\delta \textrm{m}}}$ and the four cooling collapse operators. The implementation of the scheme in an ion trap comprises of a co-trapped ``logical'' molecular ion and ``cooling'' atomic ion for which an efficient closed cycling transition exists for dissipation. The preservation of the population in the states comprising the codewords is evident under the full DEC protocol in Appendix~\ref{sec:append:dec:fulldec}.  

\paragraph*{\label{sec:implement:dec:numsim}Numerical simulation}

We simulate the dynamics of this protocol under simplified conditions similar to those considered in the sequential QEC scheme. We neglect spontaneous decay and only consider the BBR collapse operators. We consider a generic system and fix ${\Gamma_{\textrm{C}} = 1}$ with  ${\gamma_J^{\textrm{BBR}} = \gamma^{\textrm{BBR}} = 1}$. We consider the initial states~${\hat{\rho}_0 = |\overline{0}\rangle \langle \overline{0}|}$ and ${\hat{\rho}_+ = |\overline{+}\rangle \langle \overline{+}|}$ and simulate the evolution of these states over ${t = [0, 2 \, \Gamma_{\textrm{C}}^{-1}]}$.

We utilize the {CS(7,2,5)} code with the Hilbert space truncated at ${J_{\text{max}} = 10}$. This cutoff must be finite, and kept relatively small to save resources in the numerical simulation, but nonetheless should be large enough to not immediately reflect population which leaks out beyond $J_{\textrm{C}}$. All initial states are defined by their molecular rotational component and are initialized in the motional ground state for all modes.  
We examine the performance of the dissipative $J$-repumping scheme in the presence of BBR-type interactions for repumping rates ${\Omega_{\downarrow} = \Omega_{\uparrow} = 1000\, \Gamma_{\textrm{C}}}$ and cooling rates ${\Gamma_{\downarrow}^{\textrm{cool}} = \Gamma_{\uparrow}^{\textrm{cool}} = 2\ \Omega_{\downarrow,\uparrow}}$. A higher repumping rate reduces the rate of leakage.
 
\begin{figure}
    \centering
    \includegraphics[width=86.78 mm]{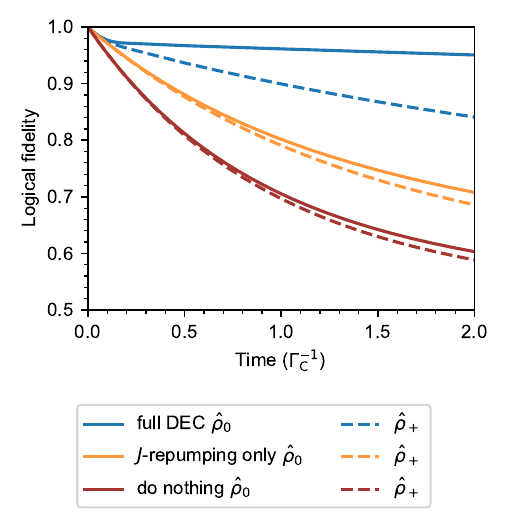}
    \caption{Evolution of the logical fidelities after tracing out the motional modes for two initial logical states comparing the full DEC scheme~(blue) to the $J$-repumping only scheme~(orange) to doing nothing~(red).}
    \label{fig:bbr-correction-fid-inf}
\end{figure}

The logical fidelities are improved with the dissipative repumping scheme over doing nothing as shown in Fig.~\ref{fig:bbr-correction-fid-inf}, although the logical fidelities deteriorate due to eventual leakage out of the codespace and correctable error subspace. The $\mathcal{F}_0$ fidelity evolution of an initial state $\hat{\rho}_0$ is in general better than that of the $\mathcal{F}_+$ fidelity evolution of an initial state $\hat{\rho}_+$. This is because the $\overline{X}$ operator is sensitive to the logical density operator coherences, whereas the $\overline{Z}$ operator is only sensitive to the populations, and over time the system decoheres due to the inability to perform amplitude refreshment. 
 
We then turn on the Zeeman correction to realize the full DEC scheme. For this, we choose the Zeeman correction rate ${\Omega_{\rightarrow,\leftarrow} = \Omega_{\uparrow,\downarrow} / 100}$. The cooling rates for the Zeeman correction modes are set to be a factor of 2 faster than the Zeeman correction Rabi rates. 

The evolution of the logical fidelities after tracing out the motional modes is shown in Fig.~\ref{fig:bbr-correction-fid-inf}. There is a clear gain in the logical fidelities with the full DEC scheme over $J$-repumping only because the population rarely leaks out of the codespace. The gain in the $\mathcal{F}_0$ fidelity of an initial state $\hat{\rho}_0$ is more pronounced over that of the $\mathcal{F}_+$ fidelity of an initial state $\hat{\rho}_+$ than in the $J$-repumping-only scheme. This is because the Zeeman DEC protocol can nearly perfectly stabilize the population in the codespace, but the coherences in the $\overline{X}$-basis still drop due to the aforementioned rotations in the amplitudes which manifest as logical errors.

\section{\label{sec:conclusion}Conclusion and outlook}

We have developed implementation strategies for exact and approximate Æ~codes in the rotation of a linear rotor within the framework of trapped ion systems employing quantum logic spectroscopy for error syndrome readout. We verified that these approaches can improve the logical fidelity evolution of a logical quantum state against decoherence from noise modeled as unresolved direct rotational transitions. The main findings can be summarized as follows:
\begin{enumerate}
    \item A hierarchy of timescales has to be fulfilled in order to minimize the extent to which the logical information leaks into the environment.
    \item An approximate encoding with sufficient distance between the Zeeman sublevels can perform better than an exact code with amplitude refreshment under sequential correction. 
    \item Dissipative error correction can perform similarly to the sequential error correction approach. It is attractive as an autonomous, non-measurement-based approach, which removes the need for classical processing and error syndrome readout and can thus be performed at a higher rate.
    \item The presented strategies can be implemented using trapped ion systems, where a molecular ion can be coupled to a co-trapped atomic ion for error syndrome or state readout via quantum logic spectroscopy. 
\end{enumerate}

Finding a molecular species that is well suited to implement the outlined strategies is a natural next step towards an experimental demonstration. Polar molecules are natural candidates as they allow ${|\delta J| = 1}$ transitions induced by BBR and spontaneous decay, and the rotation can be controlled via electric dipole transitions. A heteronuclear diatomic molecule seems to be a good candidate as it has the fewest vibrational modes. Additionally, it has only one rotational degree of freedom. The simplest system which can host Æ codes in rotational DOF without any other contributions to angular momentum is a molecule with nuclear and electronic spin singlet states. Such a molecule is in the unresolved ${|\delta J| = 1}$ regime as the transitions between different $m$-sublevels in each $J$-manifold are degenerate and thus approximates an ideal linear rotor.

There exist many candidate neutral molecular species with no nuclear or electronic spin in the ground state. However, it is not possible for singly-charged molecular ions to simultaneously possess zero nuclear and zero electronic spin, as these conditions are mutually exclusive. These conditions can be met in positive doubly-ionized diatomic molecules, known as diatomic dications (DIDIs). Polar DIDIs, such as {CaSi$^{2+}$}, which can fulfill the zero nuclear and electronic spin conditions and have a charge-to-mass ratio suitable for co-trapping with an atomic ion could host these codes without any complications from additional sources of angular momentum~\cite{Alves11}. However, many DIDIs are metastable molecules and dissociate on the second timescale or shorter and thus seem not to be suitable for QLS experiments~\cite{Sabzyan14}.

On the other hand, these codes can be adapted to allow relaxing the zero nuclear spin requirement. By allowing nuclear spin~1/2, many singly-ionized molecules with electron spin multiplicity~1 can be found. Conversely, requiring zero nuclear spin but relaxing the constraint on the electron spin multiplicity, combinations of nuclear spin~zero elements in the even parity groups of the periodic table are possible. Implementing these codes in such systems requires further research, but we are hopeful that the codes can be adapted by either operating in the Paschen-Back regime~\cite{Chou2017}, where the rotation and spin decouple, or modifying the correction pulses to also ``refresh'' the quantum number encoding the spin characteristics. The latter approach is the focus of our future research.

Our analysis shows that implementing Æ~codes and error correction in the rotation of polar molecular ions seems possible. Extending these simulations to molecular systems with more complex Hamiltonians and realistic noise models for spontaneous decay and thermal radiation is required to prove their practical feasibility. Thus, future research should focus on the analysis of other sources of noise that can contribute to decoherence. Such sources of noise include magnetic and electric field fluctuations, collisions, quantum logic infidelity due to motional heating and logic ion spontaneous decay, Stark shifts, and noisy operations~\cite{RevModPhys.91.035005}. The full time dependence of the interaction Hamiltonians describing radiation-molecule coupling, where counter-rotating terms are included, is also relevant for some processes such as far-detuned Raman and thus is important to consider in certain systems. Experimental milestones in the degree of quantum control needed to realize such QEC codes have recently been achieved in molecules and atoms. State preparation, state-resolved readout, and coherent control of {CaH$^+$} in ${J\leq 3}$, utilizing QLS via a co-trapped {Ca$^+$} logic ion with microwave carrier and CW Raman sideband transitions has been demonstrated~\cite{liu2023quantum}.

We assumed that the trapped ion system possesses 4 motional modes for the dissipative protocol, which for a string of ${N \geq 2}$ ions can be realized with certain combinations of the $N$ axial or the $2N$ radial modes. If only a single mode is available for the DEC protocol, significantly different Rabi rates or sequencing could be employed, but it is important to avoid coherent population trapping.
Furthermore, the fact that it is difficult to drive sideband transitions on rotational transitions in the microwave to terahertz regime needs to be considered in future analysis. One solution is to replace direct sideband transitions which couple a pair of manifolds with a microwave carrier transition coupling from one of the levels to an ancillary level, and a Raman ${|\delta J| = 2}$ sideband transition completing the intended coupling. This results in the desired change in the rotational and motional state with a larger Lamb-Dicke factor $\eta$ possible in the infrared or optical regime. 

On the other hand, complete control of the rotational state might even be possible in the unresolved regime. This can be achieved by using three distinct and pure polarized resonant electromagnetic fields~\cite{Leibscher2022,Pozzoli_2022}. Such control sequences can have their own challenges due to the number of required microwave operations, but could be attractive as it is not required to lift degeneracy in the rotational sublevels.

The Æ~codes considered here can be cast in terms of dephasing codes, discussed in Jain, et al.~\cite{jain2023ae} Recent theoretical work has extended these code families to more efficient encodings and has uncovered links to spin codes and binomial codes~\cite{aydin2024classcodescorrectingabsorptions}. Investigation of the feasibility and performance of these dephasing codes against realistic dephasing noise, as well as hypothetical collision codes which protect against angular momentum transfers due to collisions with background gas, will be explored in future work.

Ultimately, applications in quantum sensing or quantum computing utilizing polar molecules could benefit from employing these encoding and correction strategies. Demonstration of the improvement in logical fidelity via QEC protocols over bare encoding of quantum information in molecular systems provides the community with more options in the search for controllable yet robust quantum platforms. Future theoretical work towards fault-tolerant operations, including one-qubit rotation and 2-qubit entangling gates, is required in order to design a large-scale QC architecture based on the rotation of trapped molecules.

\stoptoc

\section*{Acknowledgements}
This research was funded by FWF 1000 Ideas project TAI-798 and ERC Horizon 2020 project ERC-2020-STG 948893. The authors thank Benjamin Stickler for reviewing the manuscript and discussions on system dynamics, Christopher Reilly for discussions on implementation and error models, and Christian Marciniak for discussions regarding calculations. The authors also acknowledge Victor Albert and Chin-wen Chou for discussions on Æ codes and error mitigation, Christiane Koch for discussion of coherent control, Florentin Reiter for discussions on dissipation engineering, Kenneth Brown for discussions on molecular ions, and the Quantum Optics \& Spectroscopy Group and associated ion trapping groups at the Universität Innsbruck for general assistance.

\textit{\textbf{Author Contributions}} 
B.F., Z.W., and P.S. developed the implementation protocols. In particular, B.F. focused on the DEC protocol, P.S. focused on the sequential QEC protocol, and Z.W. assisted with both. B.F. and P.S. performed the numerical simulations, analyzed the results, and contributed to the manuscript. Z.W., M.I.M., and S.W. contributed to discussions on implementation. All authors reviewed the manuscript. P.S. supervised the project.

\section*{Conflict of Interest}
The authors have no conflicts of interest to disclose.

\section*{Data Availability}
The source code for the simulations that support the findings of this study are openly available at \href{https://doi.org/10.5281/zenodo.14536088}{https://doi.org/10.5281/zenodo.14536088}.

\appendix

\section*{Appendix}

\resumetoc
\addtocontents{toc}{\protect\setcounter{tocdepth}{1}}
\section{\label{sec:append:rottrans}Rotational transitions}

\subsection{\label{sec:rottrans:matrixelements}Rotational part of transition dipole moment}

The Slater integrals are the integrals over the product of three spherical harmonics defined as follows, and are related to products of Clebsch-Gordon coefficients or Wigner 3-j symbols by 
\begin{equation}
    \label{eq:slaterall}
    \begin{split}
        c^{J_1}(J_3,m_3,J_2,m_2) &= \iint Y^*_{J_3,m_3}(\theta,\phi) Y_{J_2,m_2}(\theta, \phi) Y_{J_1, m_1}(\theta,\phi)\ \sin{\theta}\ \mathrm{d}\theta\ \mathrm{d}\phi\\
        &= \sqrt{\frac{(2 J_3 + 1) (2 J_2 + 1)}{4 \pi (2 J_1 + 1)}} C^{J_1, 0}_{J_3,0; J_2,0} C^{J_1, m_1}_{J_3, m_3; J_2, m_2}\\
        &= \sqrt{\frac{(2 J_1 + 1)(2 J_2 + 1)(2 J_3 + 1)}{4 \pi}} (-1)^{m_3}\\       &\quad \quad \begin{pmatrix} J_3 & J_2 & J_1 \\ -m_3 & m_2 & m_1\end{pmatrix}\begin{pmatrix} J_3 & J_2 & J_1 \\ 0 & 0 & 0\end{pmatrix}
    \end{split}
\end{equation}
The Clebsch-Gordon coefficients are defined as ${C^{J_3,m_3}_{J_1,m_1;J_2,m_2} = \langle J_1, m_1; J_2, m_2 | J_3, m_3\rangle}$ and it should be noted that $\hat{\epsilon}\cdot \hat{r} = \sqrt{4 \pi / 3}\, \big[\epsilon_z Y_{1,0}(\theta, \phi) + \epsilon_+ Y_{1,1}(\theta, \phi) + \epsilon_- Y_{1,-1}(\theta, \phi)\big]$.

\subsection{\label{sec:rottrans:einstein}Einstein coefficients and decay rates}

All rates in this work are computed in the dipole approximation. The spontaneous (SD) rate between states~${|i\rangle = |J,m\rangle \rightarrow |j\rangle = |J - 1,m + \delta m\rangle}$ is 
\begin{equation}
    \label{eq:sdrate}
    \begin{split}
    \Gamma^{\textrm{SD}}(J,m,\delta m) &= A_{ij}\\
    &= \frac{\omega_{ij}^3 |d_{ij}|^2}{3 \pi \epsilon_0 \hbar  c^3}\\
    &=\gamma^{\textrm{SD}}_J \bigg(\frac{4 \pi}{3}\bigg)|c^J(J-1,m + \delta m, 1, \delta m)|^2\, ,
    \end{split}
\end{equation}
where $A_{ij}$ is the Einstein $A$ coefficient, $d_{ij}$ is the transition dipole moment, $\omega_{ij}$ is the transition frequency between these two states, $h$ is the Planck constant, $\epsilon_0$ is the permittivity of free space, and $c$ is the speed of light. 

In this description, we separate the ${m,\delta m}$-dependent part of the transition rate due to the transition dipole moment as a coupling strength given by the Slater integral, and the ${m,\delta m}$-independent part, given by
\begin{equation}
    \label{eq:sdratesimple}
    \gamma_{J}^{\textrm{SD}} = \frac{8 d^2 B_{\textrm{R}}^3 J^3}{3\pi \epsilon_0 \hbar^4 c^3}\, ,
\end{equation}
where we assume the frequency differences between the sublevels in each manifold are small such that the rate is only associated with the  frequency difference~${\omega_{ij} = \omega_{J,J-1} = 2 B_{\textrm{R}} J / \hbar}$ between the centroids of the rotational manifolds $J$ and ${J-1}$. 

The transition dipole moment is given by
\begin{equation}
    \label{eq:dmte}
    \hat{d}_{ij} = d \langle i | \hat{r} | j \rangle
\end{equation}
with $d$ the electric dipole moment and $\hat{r}$ the unit position operator. The projection of this onto the driving electric field polarization is
\begin{equation}
    \label{eq:dmteproj}
    \begin{split}
    \hat{\epsilon}_{\delta m} \cdot \hat{d}_{ij} &= d \langle i | \hat{\epsilon}_{\delta m} \cdot \hat{r} | j \rangle\\
       &= d\, \sqrt{\frac{4\pi}{3}}\, c^{J_1}(J_2,m_2,1,\delta m)\\
       &= d\, \sqrt{\frac{4\pi}{3}} \iint Y^*_{J_2,m_2}(\theta,\phi) Y_{1,\delta m}(\theta, \phi) Y_{J_1, m_1}(\theta,\phi)\ \sin{\theta}\ \mathrm{d}\theta\ \mathrm{d}\phi\\
       &= d\, \sqrt{(2 J_1 + 1)(2 J_2 + 1)}\, (-1)^{m_2}\begin{pmatrix} J_2 & 1 & J_1 \\ -m_2 & \delta m & m_1\end{pmatrix}\begin{pmatrix} J_2 & 1 & J_1 \\ 0 & 0 & 0\end{pmatrix},
    \end{split}
\end{equation}
where ${Y_{J,m}(\theta,\phi)}$ are the spherical harmonics and ${(:::)}$ are the Wigner~{3-j}~ symbols. 

The BBR rate coupling the states ${|i\rangle = |J,m\rangle \leftrightarrow |j\rangle = |J -1, m + \delta m\rangle}$ is
\begin{equation}
    \label{eq:bbrrate}
    \begin{split}
    \Gamma^{\textrm{BBR}}(J,\delta m) &= B_{ij}\rho(\omega_{ij},T)\\
    &= \frac{\omega_{ij}^3 |d_{ij}|^2}{3 \pi \epsilon_0 \hbar c^3} \frac{1}{e^{\hbar \omega_{ij} / k_{\textrm{B}} T} - 1}\\
    &= \gamma^{\textrm{BBR}}_J \bigg(\frac{4 \pi}{3}\bigg) |c^J(J - 1,m + \delta m, 1, \delta m)|^2\, ,
    \end{split}
\end{equation}
where 
\begin{equation}
\label{eq:einsteinb}
    B_{ij} = \frac{\pi^2 c^3}{\hbar \omega_{ij}^3} A_{ij}
\end{equation}
is the Einstein $B$ coefficient, 
\begin{equation}
    \label{eq:bbrnrg}
    \rho(\omega,T) = \frac{\hbar \omega^3}{\pi^2 c^3} \frac{1}{e^{\hbar \omega / k_{\textrm{B}} T} - 1}
\end{equation}
is the spectral energy density of an environment treated as a black body thermal bath (\textit{i.e.}, assumed to follow the Planck distribution), $k_{\textrm{B}}$ is the Stefan-Boltzmann constant, and $T$ is the temperature of the environment. We assume the broadband limit, \textit{i.e.}, that the BBR energy density spectrum is broad compared to the linewidth of the rotational transition and thus we consider the pairs of states $i$ and $j$ as interacting with a single, resonant mode of frequency $\omega_{ij}$~\cite{hilborn2002einstein, milonni1988lasers, craig1998molecular}. The ${m, \delta m}$-independent part of the rates are given by
\begin{equation}
    \label{eq:bbrratesimple}
    \gamma_{J}^{\textrm{BBR}} = \frac{\gamma_{J}^{\textrm{SD}}}{e^{2 B_{\textrm{R}} J / k_{\textrm{B}} T} - 1}\, ,
\end{equation}
where ${\omega_{ij} = 2 B_{\textrm{R}} J / \hbar}$ for a linear rigid rotor.

Interactions with the environment via electric dipole transitions caused by an emission event due to spontaneous decay (SD) are given by
\begin{equation}
    \label{eq:resolvedjumpop}
    \hat{\mathcal{E}}^{\textrm{SD}}(J,m,\delta m) = \sqrt{\Gamma^{\textrm{SD}}(J,m,\delta m)}\, \rotop(J, m, -1, \delta m)
\end{equation}
which occur with rates 
\begin{equation}
    \label{eq:resolved-sd-rate}
    \begin{split}
    \Gamma^{\textrm{SD}}(J,m,\delta m) &= A_{ij}\\
    &= \gamma^{\textrm{SD}}_J \bigg(\frac{4 \pi}{3}\bigg)|c^J(J-1,m + \delta m, 1, \delta m)|^2\, ,
    \end{split}
\end{equation}
where $A_{ij}$ is the Einstein $A$ coefficient for decays between states ${|i\rangle = |J,m\rangle}$ to $|j\rangle = |J-1, m+\delta m,\delta m\rangle$ and ${c^J(J-1,m+\delta m,1,\delta m)}$ is the Slater integral.

In this description, we separate the ${m,\delta m}$-dependent part of the transition rate due to the transition dipole moment as a coupling strength given by the Slater integral, and the ${m,\delta m}$-independent part, $\gamma_{J}^{\textrm{SD}}$.

In the unresolved regime, the collapse operators which describe spontaneous decay from state $J$ are
\begin{align}
    \label{eq:sdop}
    \hat{E}^{\textrm{SD}}(J,\delta m) = \sum_{m = -J}^J \hat{\mathcal{E}}^{\textrm{SD}}(J,m,\delta m)\, .
\end{align}

\section{\label{sec:append:sequentialqec}Effect of an error on rotational state populations}

An error alters the relative amplitudes of the ${|J_{\textrm{C}},m \rangle}$~states that encode the logical qubit according to the corresponding Slater integrals. After correction, the action on the relative amplitudes is described by the rotation given in Eq.~\ref{eq:unitary-refresh-err}. The rotation angles are given by
\begin{equation}
    \label{eq:amp-refresh-angles-err}
    \theta_{\delta J, \delta m, k} = 2 \tan^{-1}\bigg[\frac{m_1 m_2 (x_k - y_k)}{x_k m_2^2 + y_k m_1^2}\bigg]\, ,
\end{equation}
where 
\begin{equation}
    \label{eq:x-refresh}
    (x_k, y_k) = \begin{cases}
        (a_{\delta J, \delta m},b_{\delta J, \delta m})\ \text{for}\ k=1\\
        (d_{\delta J, \delta m},c_{\delta J, \delta m})\ \text{for}\ k=2\, .
    \end{cases}
\end{equation}

The elements ${a_{\delta J, \delta m}, b_{\delta J, \delta m}, c_{\delta J, \delta m}, d_{\delta J, \delta m}}$ can be derived as follows: The amplitudes for the states in the logical codewords after undergoing decays with ${|\delta J| = 1}$ and subsequent equal coupling correction of $J$ and $m$ change by
\begin{equation}
    \label{eq:qmat}
    \hat{Q}_{\delta J, \delta m} = \frac{1}{n_{\delta J, \delta m}} \begin{pmatrix}
       a_{\delta J, \delta m} & 0 & 0 & 0\\
        0 & b_{\delta J, \delta m} & 0 & 0\\
        0 & 0 & c_{\delta J, \delta m} & 0\\
        0 & 0 & 0 & d_{\delta J, \delta m}
    \end{pmatrix}
\end{equation}
in the basis ${\hat{e}_1 = |J_{\textrm{C}}, -m_1\rangle}$, ${\hat{e}_2 = |J_{\textrm{C}}, m_2\rangle}$, ${\hat{e}_3 = |J_{\textrm{C}}, -m_2\rangle}$, ${\hat{e}_4 = |J_{\textrm{C}}, m_1\rangle}$. The elements of $\hat{Q}$ are
\begin{align}
    \label{eq:amplitudechanges_a}
    a_{\delta J, \delta m} &= |c^{J_{\textrm{C}}}(J_{\textrm{C}} + \delta J,-m_1 + \delta m, 1, \delta m)|\\
    \label{eq:amplitudechanges_b}
    b_{\delta J, \delta m} &= |c^{J_{\textrm{C}}}(J_{\textrm{C}} + \delta J,m_2 + \delta m, 1, \delta m)|\\
    \label{eq:amplitudechanges_c}
    c_{\delta J, \delta m} &= |c^{J_{\textrm{C}}}(J_{\textrm{C}} + \delta J,-m_2 + \delta m, 1, \delta m)|\\
    \label{eq:amplitudechanges_d}
    d_{\delta J, \delta m} &= |c^{J_{\textrm{C}}}(J_{\textrm{C}} + \delta J,m_1 + \delta m, 1, \delta m)|
\end{align}
and the norm is
\begin{equation}
    \label{eq:amp_refresh_norm}
    \begin{split}
        n_{\delta J, \delta m} &= \sqrt{\frac{m_2 a_{\delta J, \delta m}^2 + m_1 b_{\delta J, \delta m}^2}{m_1 + m_2}}\\
        &= \sqrt{\frac{m_1 c_{\delta J, \delta m}^2 + m_2 d_{\delta J, \delta m}^2}{m_1 + m_2}}\, .
    \end{split}
\end{equation} 

\section{\label{sec:append:dec}Dissipative QEC}

Simulation of the dissipative QEC protocol is performed in QuTiP using the Lindblad master equation solver. The Lindblad master equation is
\begin{equation}
    \label{eq:lindbladme}
    \frac{\textrm{d}\hat{\rho}(t)}{\textrm{d}t} = - \frac{i}{\hbar} \big[\hat{H},\hat{\rho}(t)\big] + \sum_n \frac{1}{2} \big[2 \hat{C}_n \hat{\rho}(t) \hat{C}_n^{\dagger} - \hat{\rho}(t)\hat{C}_n^{\dagger} \hat{C}_n - \hat{C}_n^{\dagger} \hat{C}_n \hat{\rho}(t)\big]
\end{equation}
where ${\hat{\rho}(t)}$ is the density operator representing the evolution of the quantum state, $\hat{H}$ is the system Hamiltonian, ${\hat{C}_n = \sqrt{\Gamma_n}\, \hat{A}_n}$ are collapse operators, $\hat{A}_n$ are the jump operators through which the environment couples to the system, and $\Gamma_n$ are the corresponding decay rates~\cite{JOHANSSON20121760,JOHANSSON20131234}. In this work, the jump operators have the form ${\rotop(J,m,\delta J, \delta m)}$ for ${|\delta J| = 1}$ and ${|\delta m| \leq 1}$. Here we set ${\hbar = 1}$ and express the Hamiltonian as an angular frequency $\Omega$ and the decay rates as ${1/e}$ rates $\Gamma_n$, which are related to the natural linewidths in angular frequency ${\Delta \omega_n = \Gamma_n}$.

\subsection{\label{sec:append:dec:decj}Dissipative repumping of rotational manifold}

To demonstrate the performance of the $J$-repumping scheme, we simulate the evolution of the system without any coupling to BBR. We consider the initial state~$\hat{\rho}_{\uparrow} (\hat{\rho}_{\downarrow}) = {\left|\uparrow \right\rangle \left\langle \uparrow \right| \big(\left|\downarrow \right\rangle \left\langle \downarrow \right|\big)}$, where $\left|\uparrow \right\rangle (\left|\downarrow \right\rangle) = {\rotop\big(7,m,+1 (-1),0\big)\,|\overline{0}\rangle}$ which represents a rotational ladder operator applied to the ${|\overline{0}\rangle}$ codeword. Upon application of the $J$-repumping scheme with the same parameters as defined in Sec.~\ref{sec:implement:dec}, the density operator is brought back toward ${\hat{\rho}_0 = |\overline{0}\rangle \langle \overline{0}|}$. The expectation value ${\langle J \rangle}$ returns to ${J_{\textrm{C}} = 7}$ for either initial error state as expected, shown in Fig.~\ref{fig:simple-correction-1-j}.

The evolution of the physical fidelity ${F\big(\hat{\rho}_0,\hat{\rho}(t) \big) = \textrm{tr}\big(\hat{\rho}(t) \hat{\rho}_0\big)}$, after tracing out the motional modes, is shown in Fig.~\ref{fig:simple-correction-1-physfid}. Upon application of the $J$-repumping scheme with the same parameters as defined in Sec.~\ref{sec:implement:dec}, the density is brought back toward $\hat{\rho}_0$. At ${t = 0.05\, \Gamma_{\textrm{C}}^{-1}}$, the physical fidelities reach 0.974 for an initial state $\hat{\rho}_{\downarrow}$ and 0.987 for an initial state~${\hat{\rho}_{\uparrow}}$. The downward correction of $\hat{\rho}_{\uparrow}$ from ${J = 8}$ reaches a higher fidelity than the upward correction of $\hat{\rho}_{\downarrow}$ from ${J = 6}$ in this time due to the slightly different effective repumping rates arising from the different Slater integrals for each coupling. 

\subsection{\label{sec:append:dec:decz}Zeeman dissipative error correction}

\sloppy To demonstrate the performance of the full DEC scheme with rotational manifold repumping and Zeeman DEC combined, we can simulate without any coupling to BBR. We consider the initial states ${\hat{\rho}_{\leftarrow} (\hat{\rho}_{\rightarrow}) = \left|\leftarrow\rangle \langle \leftarrow \right| \big(\left|\rightarrow\rangle \langle \rightarrow \right|\big)}$, where ${\left|\leftarrow\right\rangle (\left|\rightarrow\right\rangle) = \rotop\big(7,m,0,-1 (+1)\big)\left|\overline{0}\right\rangle}$. This represents a ladder operator applied to the ${|\overline{0}\rangle}$ codeword followed by perfect repumping back to ${J_{\textrm{C}} = 7}$. The expectation value ${\langle J \rangle}$ is fixed at ${J_{\textrm{C}} = 7}$ as there is no operator which maps out of the code manifold. Upon application of the Zeeman correction scheme with the same parameters as defined in Sec.~\ref{sec:implement:dec}, the density operator is brought toward $\hat{\rho}_0$.

The evolution of the physical fidelity with respect to $\hat{\rho}_0$, after tracing out motional modes, is ${F\big( \hat{\rho}_0, \hat{\rho}(t) \big)}$ and is shown in Fig.~\ref{fig:simple-correction-2-physfid}. After a short time, the physical fidelities for both $m$-shifted states symmetrically reach a value of 0.994 and approach arbitrarily close to 1. There is a small oscillation in the fidelity due to the finite cooling rate, as some population is cycled back to the error state on the red sideband before being returned on the blue sideband and cooled further.

\subsection{\label{sec:append:dec:fulldec}Full dissipative error correction}

The comparison of the action of the full DEC scheme to $J$-repumping only to doing nothing can be observed by examining the final state populations after tracing out the motional modes. This is shown for the same parameters defined in Sec.~\ref{sec:implement:dec} in Fig.~\ref{fig:bbr-correction-2-rothinton}. The concatenation of the Zeeman DEC scheme with the $J$-repumping DEC scheme significantly improves the population preserved in $\mathcal{H}_{\textrm{C}}$.

\begin{multicols}{2}
\begin{Figure}
    \centering
    \includegraphics[width=\columnwidth]{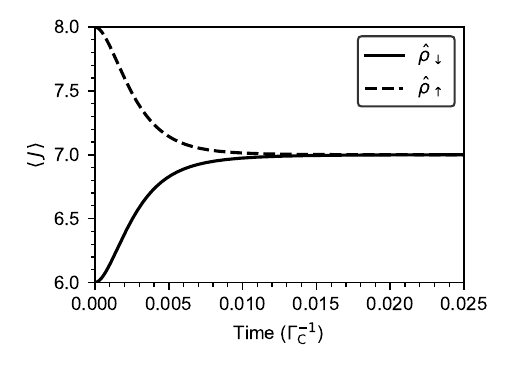}
    \captionof{figure}{Evolution of the expectation value of $J$ for two different initial error states using the dissipative $J$-repumping scheme.}
    \label{fig:simple-correction-1-j}
\end{Figure}

\begin{Figure}
    \centering
    \includegraphics[width=\columnwidth]{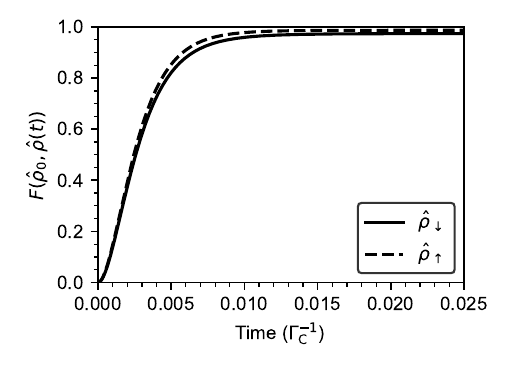}
    \captionof{figure}{Evolution of the physical fidelity ${F\big(\hat{\rho}_0,\hat{\rho}(t)\big)}$ after tracing out the motional modes for two different initial error states using the dissipative $J$-repumping scheme.}
    \label{fig:simple-correction-1-physfid}
\end{Figure}

\begin{Figure}
    \centering
     \includegraphics[width=\columnwidth]{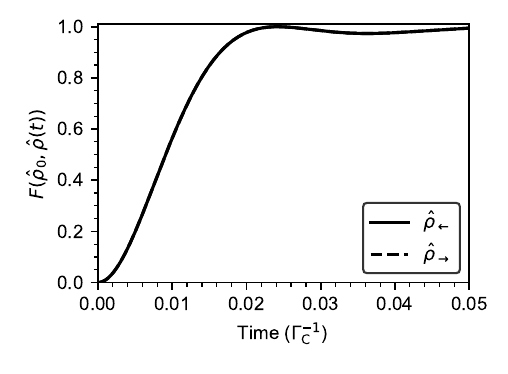}
    \captionof{figure}{Evolution of the physical fidelity ${F\big( \hat{\rho}_0,\hat{\rho}(t)\big)}$ after tracing out the motional modes for two different initial error states using the Zeeman DEC scheme.}
    \label{fig:simple-correction-2-physfid}
\end{Figure}

\begin{Figure}
    \centering
   \includegraphics[width=\columnwidth]{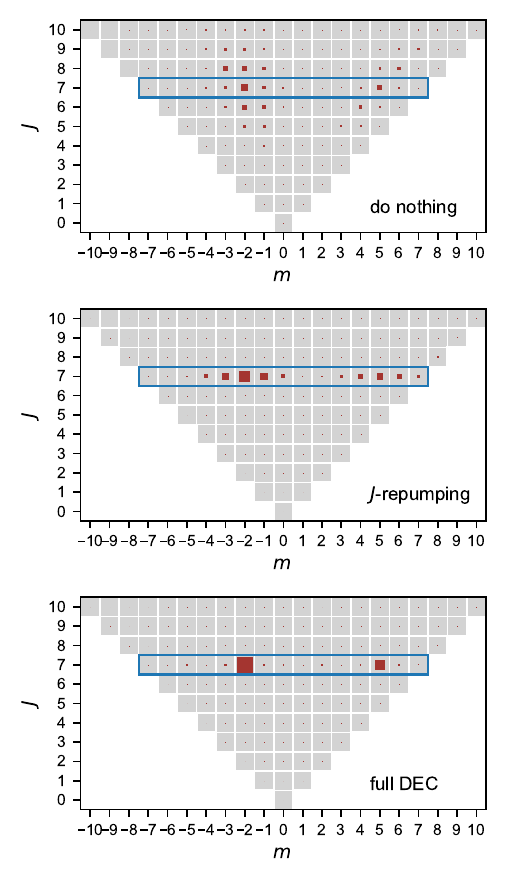}
    \captionof{figure}{Final states at ${t=2\, \Gamma_{\textrm{C}}^{-1}}$ for initial state~${\hat{\rho}_0}$ after doing nothing~(top), after $J$-repumping only~(middle), and after full DEC~(bottom) having traced out the motional modes. The populations of each state are indicated with red squares with areas proportional to the populations. DEC rate parameters are the same as those used in the main text. The code manifold ${J_{\textrm{C}} = 7}$ is outlined in blue.}
    \label{fig:bbr-correction-2-rothinton}
\end{Figure}

\clearpage
\end{multicols}
\bibliographystyle{quantum}
\bibliography{mainv2}
\end{document}